\documentclass{aa}
\usepackage{txfonts}
\usepackage{array}  
\newcolumntype{L}[1]{>{\raggedright\arraybackslash}p{#1}}
\usepackage{natbib}
\bibpunct{(}{)}{;}{a}{}{,}    
\usepackage{graphicx}
\usepackage[utf8]{inputenc}
\usepackage[T1]{fontenc}
\usepackage{textcomp}
\usepackage{subfigure}
\usepackage{multirow}

\DeclareUnicodeCharacter{0301}{\'{e}} 

\usepackage[breaklinks=true, colorlinks=true, urlcolor=magenta, citecolor=blue, linkcolor=blue]{hyperref}

\newcommand{\po}{$P_{\Omega}$\,} 
\newcommand{\ptwoo}{$P_{2\Omega}$\,} 
\newcommand{\pthreeo}{$P_{3\Omega}$\,} 
\newcommand{\pfouro}{$P_{4\Omega}$\,} 
\newcommand{\pfiveo}{$P_{5\Omega}$\,} 
\newcommand{\psixo}{$P_{6\Omega}$\,} 
\newcommand{\pseveno}{$P_{7\Omega}$\,} 
\newcommand{\peighto}{$P_{8\Omega}$\,} 
\newcommand{\pnineo}{$P_{9\Omega}$\,} 

\begin{document}

\title{Unveiling the nature of two dwarf novae: CRTS J080846.2+313106 and V416 Dra}

\author{Arti Joshi\inst{\ref{PUC} ,\ref{IIA}}\thanks{ajoshi@astro.puc.cl, aartijoshiphysics@gmail.com} \and
         M{\'a}rcio Catelan\inst{\ref{PUC}, \ref{MIA}} \and
        Simone Scaringi \inst{\ref{CEA}} \and 
        Axel Schwope\inst{\ref{AIP}} \and
        G. C. Anupama\inst{\ref{IIA}} \and
        Nikita Rawat\inst{\ref{ARIES}}\and
        Devendra K. Sahu  \inst{\ref{IIA}}\and
        Mridweeka Singh \inst{\ref{IIA}} \and 
        Raya Dastidar \inst{\ref{UNAB}, \ref{MIA}} \and
        Rama Venkata Subramanian \inst{\ref{ASP}} \and 
        Srinivas M Rao  \inst{\ref{ARIES}}        
       }
       
\institute{
Institute of Astrophysics, Pontificia Universidad Católica de Chile, Av. Vicuña MacKenna 4860, 7820436 Santiago, Chile \label{PUC}
\and
Indian Institute of Astrophysics, Koramangala, Bangalore 560 034, India \label{IIA}
\and
Millennium Institute of Astrophysics, Nuncio Monse\~{n}or S\'{o}tero Sanz 100, Providencia, Santiago, Chile \label{MIA}
\and
Centre for Extragalactic Astronomy, Department of Physics, Durham University, South Road, Durham DH1 3LE, UK \label{CEA}
\and
Leibniz-Institut für Astrophysik Potsdam, An der Sternwarte 16, 14482 Potsdam, Germany\label{AIP}
\and
Aryabhatta Research Institute of Observational Sciences, Manora Peak, Nainital 263001, India\label{ARIES}
\and
Instituto de Astrofı́sica, Universidad Andres Bello, Fernandez Concha 700, Las Condes, Santiago RM, Chile \label{UNAB}
\and
Department of Sciences, Amrita School of Physical Sciences, Amrita Vishwa Vidyapeetham, Coimbatore, TamilNadu, 641112, India \label{ASP}
}

\abstract
{%
We present the analysis of optical photometric and spectroscopic observations of two non-magnetic cataclysmic variables, namely CRTS J080846.2+313106 and V416 Dra. CRTS J080846.2+313106 has been found to vary with a period of  4.9116$\pm$0.0003 h, which was not found in earlier studies and is provisionally suggested as the orbital period of the system. In both long-period systems, the observed dominant signal at second harmonic of the orbital frequency and the orbital modulation during quiescence are suggestive of ellipsoidal variation from changing aspects of the secondary, with an additional contribution from the accretion stream or hotspot. However, during the outburst, the hotspot
itself is overwhelmed by the increased brightness, which is possibly associated with the accretion disc. The mid-eclipse phase for V416 Dra occurs earlier and the width of the eclipse is greater during outbursts compared to quiescence, suggesting an increased accretion disc radius during outbursts. Furthermore, the investigation of accretion disc eclipse in V416 Dra implies that a total disc eclipse is possible during quiescence, whereas the disc seems to be partially obscured during outbursts, which further signifies that the disc may grow in size as the outburst progresses.  
 Optical spectra of CRTS J080846.2+313106 and V416 Dra are typical of dwarf novae during quiescence, and they both  show a significant contribution from the M2-4V secondary. The light curve patterns, orbital periods, and spectra observed in both systems look remarkably similar, and seem to resemble the characteristics of U Gem-type dwarf novae.
}
\keywords{accretion, accretion discs -- stars: novae, cataclysmic variables -- stars: individual: CRTS J080846.2+313106 --- stars: individual: V416 Dra}

\titlerunning{CRTS J080846.2+313106 and V416 Dra}
\authorrunning{Joshi et~al.}
\maketitle

\begin{table*}
\small
\centering
\caption{Log of the \textit {TESS} and spectroscopic observations.\label{tab:obslog}}
\renewcommand{\arraystretch}{1}
\begin{tabular}{lccccccc}
\hline
&&  \multicolumn{4}{c}{\textit {TESS} Observations} \\
\cline{2-6}
Object  & Sector  &  Start Time &  End Time & Exp. (s) & Total Span     \\
\hline                                CRTS J080846.2+313106&  44      &2021-10-12T16:51:50.6 & 2021-11-05T22:35:13.3 & 120 & 24.2 \\
                     &  45      &2021-11-07T05:27:23.6 & 2021-12-02T03:04:24.8 & 120  & 24.9 \\
                     &  46      &2021-12-03T12:08:31.2 &2021-12-30T04:56:31.8  & 120 & 26.7 \\
                     &  47      &2021-12-31T07:40:34.6 & 2022-01-27T10:40:51.5 & 120 & 27.1 \\
                     &  71      & 2023-10-16T12:53:47.2& 2023-11-11T10:53:21.3 & 120  & 25.9  \\
                     &          & 2023-10-16T12:53:47.2& 2023-11-11T11:17:21.4 & 20 & 25.9  \\
                     &  72      & 2023-11-11T16:25:23.5& 2023-12-07T02:04:19.8 & 120 & 25.4 \\
                     &          &2023-11-11T16:25:23.5& 2023-12-07T02:05:39.8 & 20 & 25.4 \\
V416 Dra             &  14      &2019-07-18T20:27:02.5 & 2019-08-14T16:53:28.5 & 120& 26.8\\ 
                     &  15      &2019-08-15T20:43:29.8 & 2019-09-10T21:49:54.8 & 120 & 26.0 \\
                     &  16      &2019-09-12T03:37:56.1 & 2019-10-06T19:40:14.3 & 120& 24.6 \\
                     &  17      &2019-10-08T04:24:15.9 & 2019-11-02T04:40:25.2 & 120& 25.0 \\
                     &  19      &2019-11-28T14:00:23.0 & 2019-12-23T15:28:09.1 & 120& 25.0 \\
                     &  20      &2019-12-25T00:04:08.3 & 2020-01-20T07:43:44.0 & 120& 26.3 \\
                     &  21      &2020-01-21T22:23:42.5 & 2020-02-18T06:43:12.4 & 120& 27.3 \\
                     &  22      &2020-02-19T19:41:10.8 & 2020-03-17T23:48:43.1 & 120& 27.1  \\
                     &  23      &2020-03-19T14:30:41.8 & 2020-04-15T08:58:22.9 & 120& 26.7 \\
                     &  26      &2020-06-09T18:22:21.3 & 2020-07-04T15:12:36.9 & 120& 24.8 \\
                     &  40      &2021-06-25T03:38:15.6 & 2021-07-23T08:28:38.9 & 120& 28.2 \\
                     &  41      &2021-07-24T11:42:39.8 & 2021-08-20T01:53:06.3 & 120& 26.5 \\
                     &  47      &2021-12-31T07:33:34.2 & 2022-01-27T10:33:07.7 & 120& 27.1 \\
                     &  49      &2022-02-26T23:18:34.0 & 2022-03-25T19:32:08.6 & 120& 26.8 \\
                     &  50      &2022-03-26T18:28:07.7 & 2022-04-22T00:11:51.8 & 120& 26.2 \\
                     &  51      &2022-04-23T10:53:51.0 & 2022-05-18T00:47:47.3 & 120& 24.5 \\
                     &  52      &2022-05-19T03:11:46.7 & 2022-06-12T13:47:53.9 & 120& 24.4 \\
                     &  53      &2022-06-13T11:51:54.3 & 2022-07-08T11:28:11.1 & 120& 24.9 \\
                     &  54      &2022-07-09T09:32:12.3 & 2022-08-04T15:04:36.7 & 120& 26.2 \\
                     &  55      &2022-08-05T14:22:37.4 & 2022-09-01T18:19:04.6 & 120& 27.1 \\
                     &  56      &2022-09-02T18:09:05.3 & 2022-09-30T15:19:28.6 & 120& 27.8 \\
                     &  57      &2022-09-30T20:29:28.8 & 2022-10-29T14:45:41.9 & 120& 28.7 \\
                     &  58      &2022-10-29T19:53:41.9 & 2022-11-26T13:09:40.8 & 120& 27.7 \\
                     &  59      &2022-11-26T18:19:40.7 & 2022-12-23T04:29:26.7 & 120& 26.4 \\
                     &  60      &2022-12-23T09:39:26.6 & 2023-01-18T01:59:03.4 & 120& 25.6 \\
                     &  73      &2023-12-07T07:09:22.3 & 2024-01-03T03:35:03.5 & 120& 26.8 \\
                     &  73      &2023-12-07T07:09:22.2 & 2024-01-03T03:35:23.5 &  20& 26.8 \\
\hline
\end{tabular}
\setlength{\tabcolsep}{0.2in}
\centering
\centering
\begin{tabular}{@{}l  l l l l@{}}
&  \multicolumn{3}{c}{Spectroscopic Observations} \\
\cline{2-5}
Object & Telescope/Instrument & Wavelength Range & Date of Obs. & Exp. (s) \\
\hline
CRTS J080846.2+313106 & HCT/HFOSC-Gr7 & 3800-6840 \AA & 17-12-2018 & 2700 \\
                      & HCT/HFOSC-Gr7 & 3800-6840 \AA & 25-11-2022 & 2700 \\
                      & HCT/HFOSC-Gr7 & 3800-6840 \AA & 26-11-2022 & 3600 \\
V416 Dra              & HCT/HFOSC-Gr7 & 3800-6840 \AA & 21-03-2023 & 3600 \\
                      & HCT/HFOSC-Gr7 & 3800-6840 \AA & 21-04-2023 & 2620 \\
                      & HCT/HFOSC-Gr7 & 3800-6840 \AA & 20-05-2023 & 3300 \\
\hline
\end{tabular}
\end{table*}

\begin{figure*}
\centering
\subfigure[]{\includegraphics[width=\textwidth]{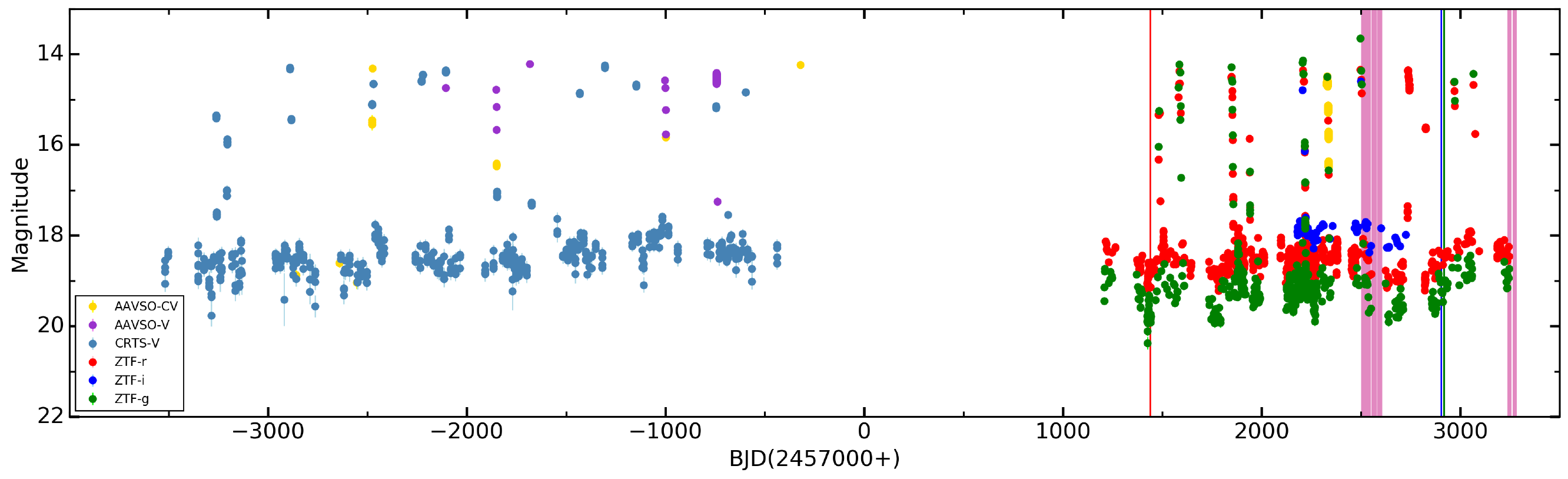}\label{fig:longtermlc_J0808}}
\subfigure[]{\includegraphics[width=\textwidth]{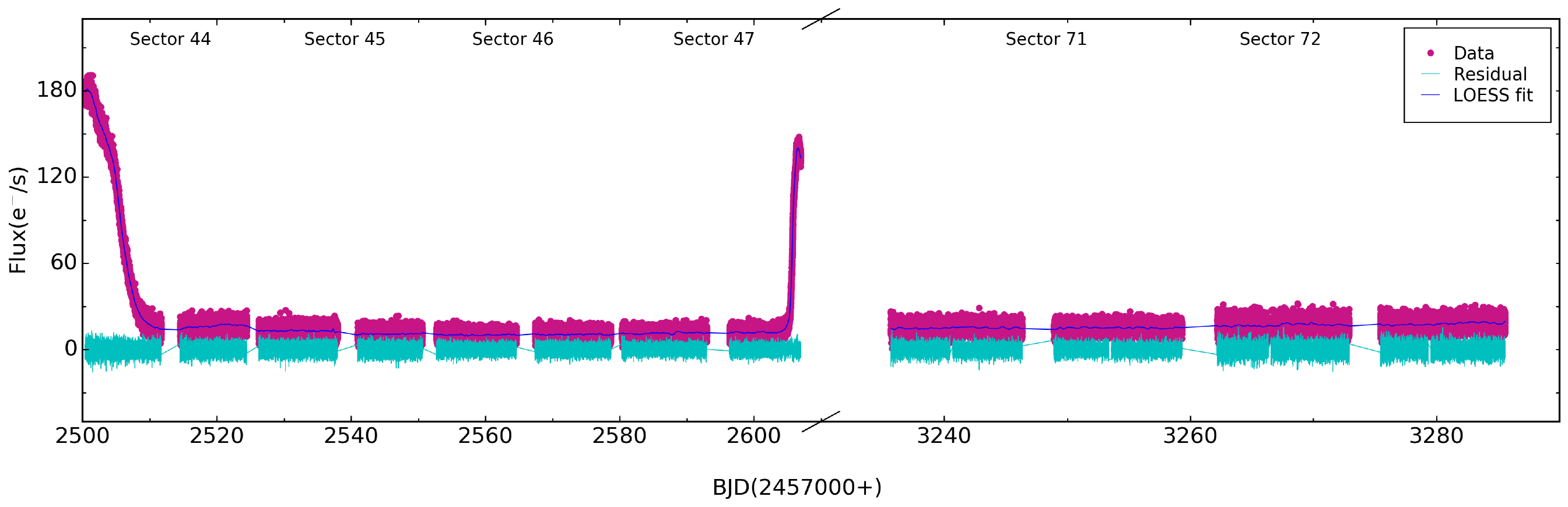}\label{fig:tesslc_J0808}}
\caption{(a) Long-term light curve of J0808, where the pink shaded regions correspond to the \textit{TESS} observations in each sector. Three vertical lines (red, blue, and green) represent the successive epoch of spectroscopic observations. (b) The zoomed-in version of the \textit {TESS} observations of J0808, displaying two outbursts. The blue solid line represents the smoothed light curve using the LOESS fit. The light green line is the detrended light curve after subtracting the smoothed light curve .} 
\end{figure*}

\begin{figure*}
\centering
\subfigure[]
{\includegraphics[width=\textwidth]{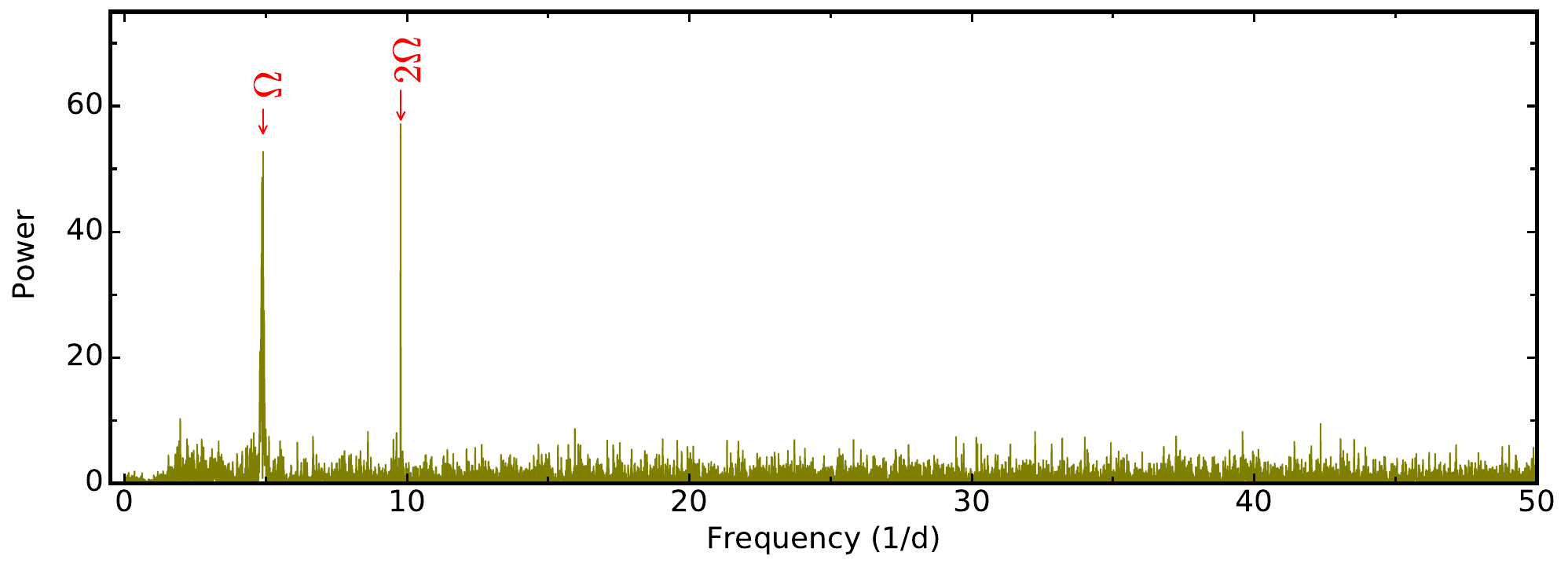}\label{fig:tessps_J0808}}
\subfigure[]
{\includegraphics[width=0.48\textwidth]{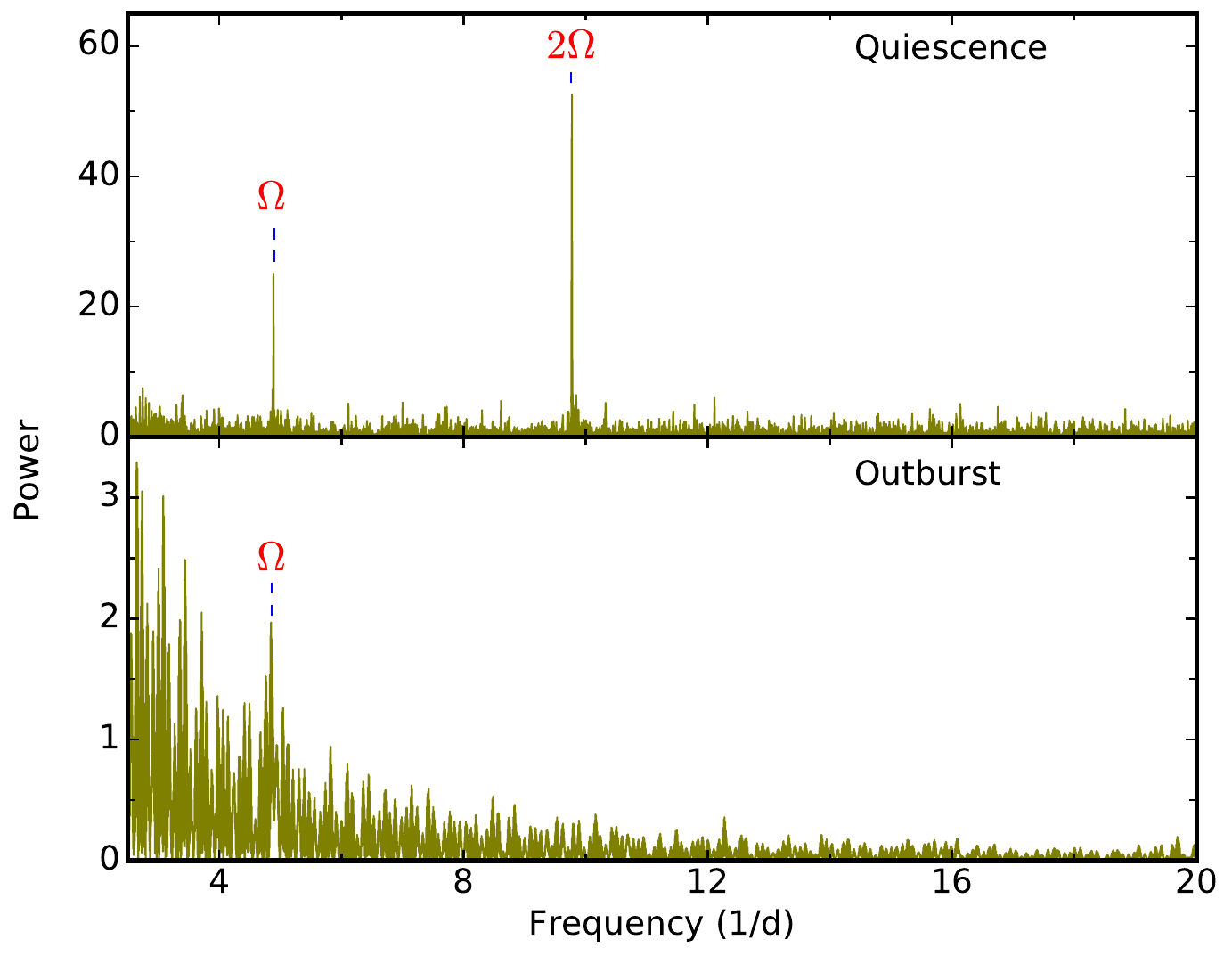}\label{fig:tessps_qs_ob_J0808}}
\subfigure[]
{\includegraphics[width=0.48\textwidth]{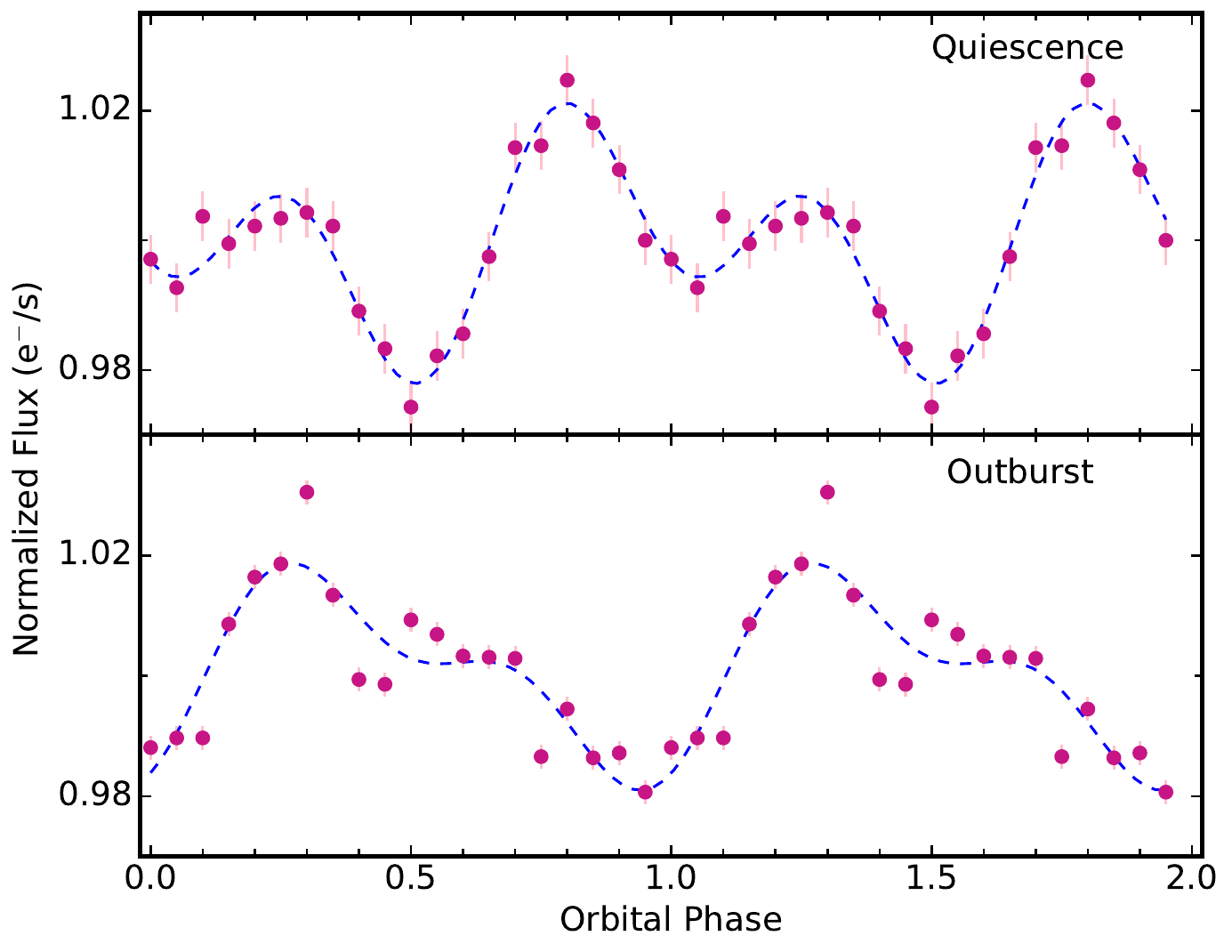}\label{fig:tessflc_qs_ob_J0808}}
\caption{(a) LS periodogram of J0808 obtained using the combined set of all sectors detrended light curves in Figure \ref{fig:tesslc_J0808}. The strongest signals corresponds to the frequencies $\Omega$ and $2\Omega$ are marked. (b)  The LS power spectra, obtained from the quiescent (from BJD 2459514.5319488 to BJD 2460285.5864469) and outburst (from BJD 2459500.4000170 to BJD 2459511.7247343 in sector 44, and from BJD 2459604.0007540 to BJD 2459606.9451476 in sector 47) time-spans, are zoomed in on interesting peaks. (c)  Orbital-phase-folded \textit {TESS} light curves obtained from the quiescent and outburst durations with phase bin of 0.05.} 
\end{figure*}

\begin{figure*}
\centering
\includegraphics[width=0.7\textwidth]{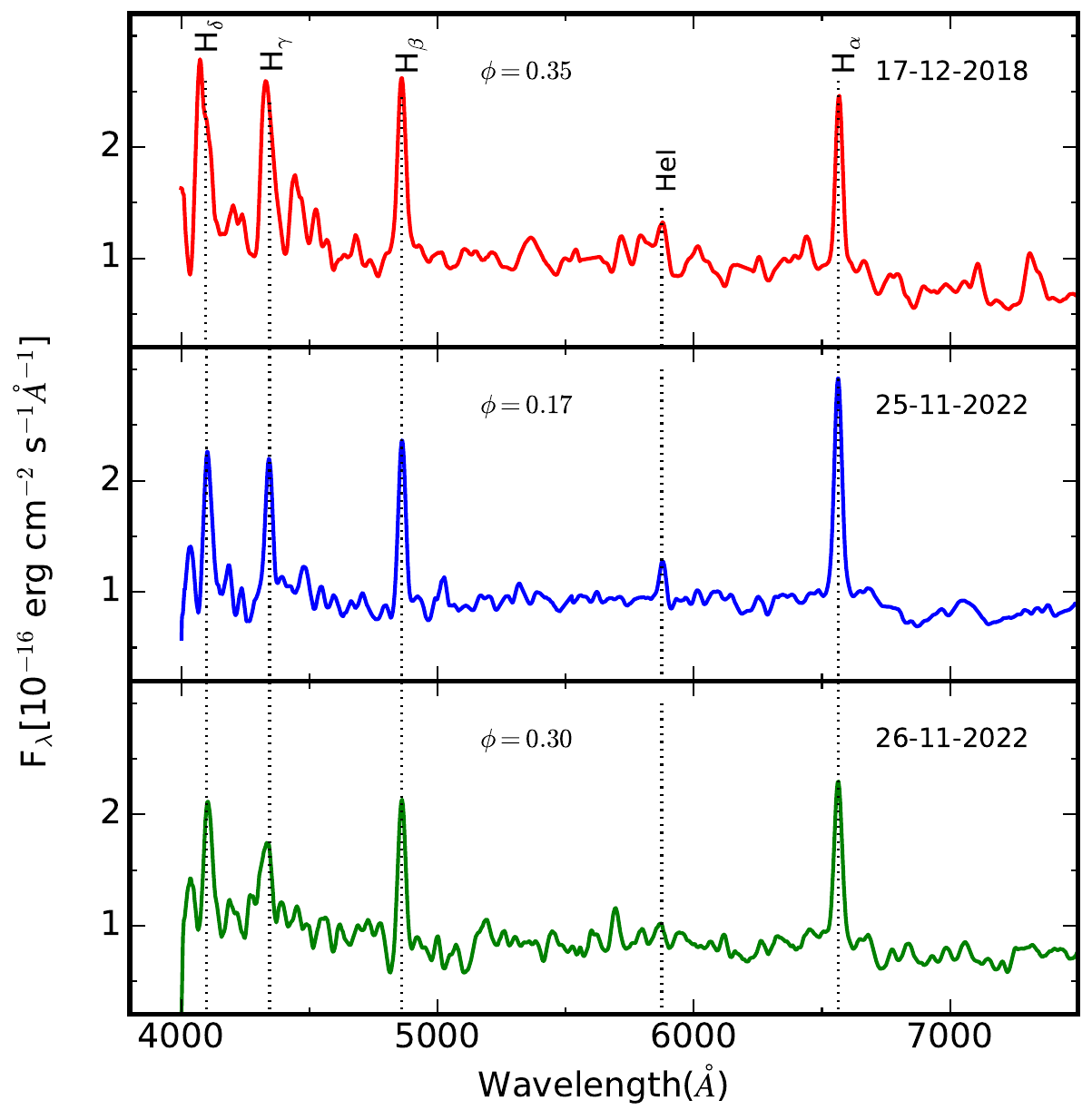}
\caption{Optical spectra of J0808 for three epochs of observations. The orbital phase and epoch of observation is mentioned in each panel.} \label{fig:spec_J08}
\end{figure*}


\section{Introduction}
\label{sec:intro}
Cataclysmic variables (CVs) are semi-detached binaries consisting of a primary white dwarf (WD) accreting material from a main sequence secondary companion \citep{Warner95}. CVs are broadly classified into two categories: non-magnetic CVs and magnetic CVs. In the absence of strong magnetic fields, the conservation of angular momentum forces accreting material to form an accretion disc around the WD, where viscous forces cause it to slowly move inwards and to finally settle on the surface of the compact object. In optical light, this disc is usually the dominant source of radiation. The accretion disc may suffer recurrent episodes of instability known as dwarf nova outbursts, which increase the luminosity of the system, typically 2 to 8 mag  over timescales ranging from days to decades. Each outburst lasts for several days to weeks. Among dwarf novae, the subtype called U Gem typically displays normal outbursts that last less than two weeks and have a typical amplitude of 2$-$5 mag \citep{Warner95}. These systems are commonly found above the 2-3 h period gap. 
 The outburst durations in U Gem sometimes exhibit a pattern where a long outburst follows a short outburst after a significant time duration. This suggests the possibility of a bimodal outburst distribution, as observed in other dwarf novae \citep{VanParadijs83, Szkody84, Ak02}. The alternation of long and short outbursts in the CVs above the period gap has been shown to be a natural consequence of the mass-transfer fluctuations \citep{Buat-menard01}. 
  On the other hand, SU UMa dwarf novae, which belong to another subtype, lie below the period gap and exhibit superoutbursts in addition to the normal outbursts, which uniquely display superhumps with periods closely aligned with the orbital period of the system  \citep{Vogt74, Warner75, Warner95, Osaki96}. Superoutbursts typically have a longer duration, lasting approximately 12 to 20 days, with a typical amplitude of around 3-5 mag. In contrast, normal outbursts last several days, with an amplitude typically around 2-3 mag.

In this paper, we selected two poorly studied dwarf novae, CRTS J080846.2+313106 (hereafter J0808) and V416 Dra, from the catalogue of \citet[][update RKcat7.24, 2016, RK\footnote{\url{https://wwwmpa.mpa-garching.mpg.de/RKcat/cbcat}}]{Ritter03} for a detailed investigation. J0808 was identified as a CV by \cite{Szkody04} based on the optical spectrum obtained by the Sloan Digital Sky Survey (SDSS) in the year 2002. The SDSS spectrum obtained during a likely dwarf nova outburst revealed the feature of broad absorption lines along with weak emissions in the core. Moreover, their 9 h photometric observations showed a modulation with an amplitude of 0.04 mag and suggested an ambiguous period of 6 h for J0808.   Later, \cite{Han20} confirmed its identification as a CV based on a Large Sky Area Multi-Object Fiber Spectroscopic Telescope \citep[LAMOST;][]{Cui12} spectrum. 
This source has a somewhat uncertain parallax. Accordingly, {\em Gaia}'s Data Release 3 (DR3) gives a distance to J0808 of $1085^{+387}_{-200}$~pc, using the Generalized Stellar Parametrizer  
 from Photometry (GSP-Phot) method \citep{Gaia-Fouesneau23}.

V416 Dra was spectroscopically identified as a dwarf nova in the Hamburg Quasar Survey \citep{Hagen95}.  \cite{Aungwerojwit06} investigated V416 Dra using photometric and spectroscopic observations. Their analysis revealed eclipsing photometric light curves, allowing them to determine the orbital period of V416 Dra as 272.317$\pm$0.001 min. They found a radial velocity amplitude of  128.0$\pm$9.6 km s$^{-1}$. A high orbital inclination was also proposed, based on the presence of double-peaked Balmer emission lines and the detection of eclipses in the light curves. In their spectroscopic observations, for a few epochs, strong emission lines were observed and for a few epochs, broad absorption troughs were seen, which is typical of a dwarf nova during quiescence and outburst. Additionally, they hinted ambiguously that V416 Dra might be a Z Cam type dwarf nova.  
 For this source, using directly the well-defined {\em Gaia} DR3 parallax \citep{Vallenari23}, we obtain a distance of $608_{+14}^{-13}$~pc.

In light of earlier studies, it is clear that the light curve morphologies, orbital period, and a proper class of J0808 are yet to be known, which is essential to probe the nature of this system. Moreover, an ambiguous classification of V416 Dra encouraged us to revisit its properties in detail. Therefore, with a motivation to ascertain their true nature, we intend to carry out detailed analyses of these two systems using the publicly available extensive photometric Transiting Exoplanet Survey Satellite \citep[\textit{TESS};][]{Ricker15} observations as well as ground-based spectroscopic observations. The paper is organized as follows. In Section \ref{sec:obs}, we summarize optical observations and data reduction description. Section \ref{sec:analysis_results} contains analyses and the results. Finally, we present the discussion and summary in Sections \ref{sec:dis} and \ref{sec:sum}, respectively.


\section{Observations and data reduction}\label{sec:obs}
\subsection{\textit {TESS}, CRTS, AAVSO, and ZTF observations}
\label{sec:obs_phot}
A detailed log of the \textit{TESS} observations for both sources is given in Table \ref{tab:obslog}. The cadence for both sources was 2 min. However, data were also available at a 20 s cadence in sectors 71 and 72 for J0808 and in sector 73 for V416 Dra. The data of J0808 and V416 Dra are stored in the Mikulski Archive for Space Telescopes (MAST) data archive\footnote{\url{https://mast.stsci.edu/portal/Mashup/Clients/Mast/Portal.html}} with identification numbers `TIC 165818951' and `TIC 229791903', respectively. 
\textit{TESS} consists of four cameras, each with a field of view of 24$\times$24 degree$^2$, which are aligned to cover 24$\times$96-degree$^2$ strips of the sky called `sectors' \citep[see][for details]{Ricker15}. \textit{TESS} bandpass extends from 600 to 1000 nm, with an effective wavelength of 800 nm. Data taken during an anomalous event had quality flags greater than 0 in the FITS file data structure, and thus we have considered only the data with the `QUALITY flag' = 0. In the case of J0808, the pre-search data-conditioned simple aperture photometry (PDCSAP) presents abrupt changes in flux for sectors 44 and 47. To inspect this unexpected behaviour of light curves, we further checked simple aperture photometry (SAP) light curves for all four sectors. We have found that the PDCSAP and SAP light curves are similar for sectors 45 and 46; however, they are different from each other for sectors 44 and 47. Thus, to examine this discrepancy, we have checked the light curve variations by downloading the target pixel files and found 
 that the brightness of the object in the target pixel file aligns with the SAP light curves of both sectors, unlike the PDCSAP light curves. Further, to determine which light curve to use, we have also compared both the SAP and PDCSAP light curves with the available simultaneous Zwicky Transient Facility \citep[ZTF\footnote{\url{https://irsa.ipac.caltech.edu/frontpage/}};][]{Bellm19, Graham19, Dekany20}
  g and r band photometry for sector 44. However, no ZTF-g/r observations are available to compare \textit {TESS} observations of sector 47. Following this approach, we found that the SAP flux of sector 44 agreed well with the ZTF data rather than the PDCSAP flux. PDSCAP and SAP flux discrepancies were also seen in V416 Dra for sectors 21, 22, 23, 26, 50, 54, 55, 57, 59, and 60. After carefully examining their target pixel files and comparing the \textit{TESS} magnitudes with the simultaneous ground-based ZTF-g/r band photometry, we found that the SAP flux values are more suitable than the PDCSAP flux values for all these sectors. Consequently, we elected to use the SAP flux values for further analysis of J0808 and V416 Dra. 

We have also utilized the V band data of J0808 from the  Catalina Schmidt Survey  \citep[CSS\footnote{\url{http://nesssi.cacr.caltech.edu/DataRelease/}};][]{Drake09, Larson03} for our analysis. Further, the optical photometric data of J0808 and V416 Dra were collected from the American Association of Variable Star Observers \citep[AAVSO\footnote{\url{https://www.aavso.org/}};][]{kafka} in the CV (unfiltered data with a V-band zero-point) and  V (Johnson V) bands to represent the long-term variability of both sources. The ZTF observations of J0808 and V416 Dra were also acquired in the ZTF-g, ZTF-r, and ZTF-i filters. 
The long-term CRTS, AAVSO, and ZTF light curves of J0808 and V416 Dra are easily accessible online.  

\subsection{Optical spectroscopy} 
\label{sec:obs_spec}
The spectroscopic observations of both sources were obtained using the 2.01-m Himalayan Chandra Telescope (HCT) at Indian Astronomical Observatory, Hanle, which is equipped with the Hanle Faint Object Spectrograph and Camera. The observing log for each observation is given in Table \ref{tab:obslog}. For all epochs, the observations were taken using a slit width of 1.92 arcsec and a Gr7 (3800-6840 \AA) grism with a resolving power of 1330. During each observing run, spectrophotometric standard stars and FeAr arc lamps were also observed for flux and wavelength calibration, respectively. Slit loss corrections were applied to the 17 December 2018 and  26 November 2022 spectra of J0808, and 21 April 2023 and 20 May 2023 spectra of V416 Dra, using the existing ZTF photometry corresponding to these epochs of spectroscopic observations. However, for the 25 November 2022 spectrum of J0808 and 21 March 2023 spectrum of V416 Dra, the spectroscopic flux matched with the ZTF-r/g photometric flux, and hence no correction was applied. Extinction correction was also applied to the spectra of each epoch using a standard extinction law with $R_{V}$=3.1 \citep{Schlafly11}.
The spectra  were extracted using standard tasks in IRAF\footnote{
IRAF was distributed by the National Optical Astronomy Observatory, which was managed by the Association of Universities for Research in Astronomy (AURA) under a cooperative agreement with the National Science Foundation} and reduced flux-calibrated spectra were used for further analysis.
 
\begin{figure*}[ht]
\centering
\includegraphics[width=\textwidth]{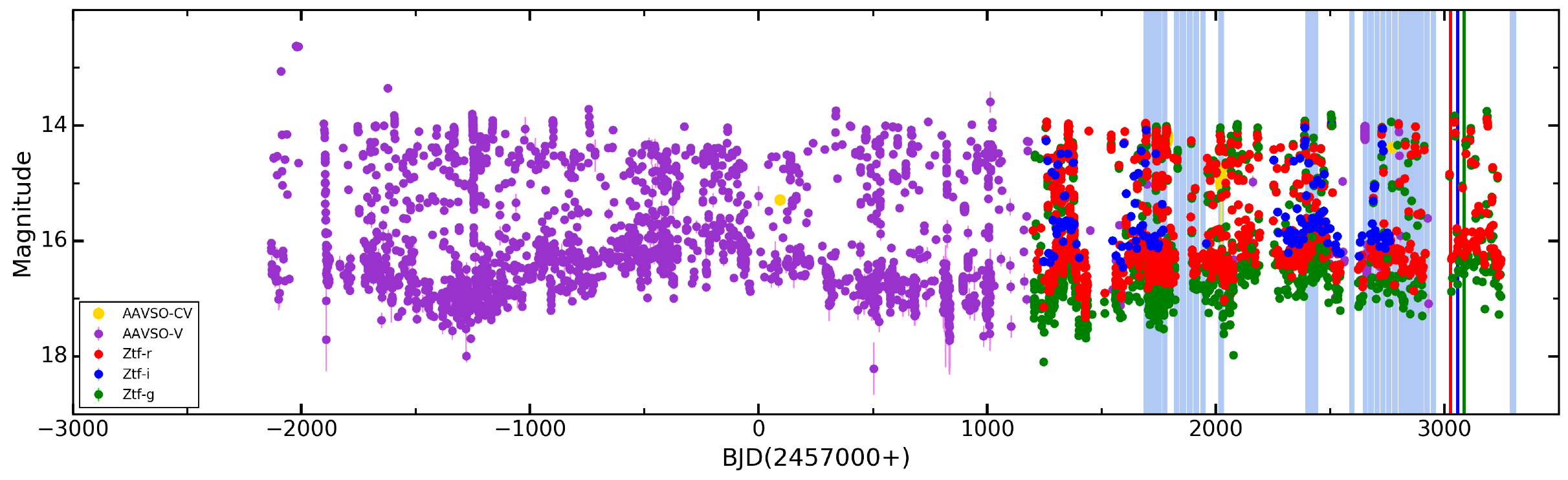}
\caption{Long-term light curve of V416 Dra, where the light blue shaded regions correspond to the \textit {TESS} observations in each sector. Three vertical lines (red, blue, and green) represent the successive epoch of spectroscopic observations.}
\label{fig:longtermlc_v416dra}
\end{figure*}

\begin{figure*}[h!]
\centering
\subfigure[]{\includegraphics[width=0.92\textwidth]{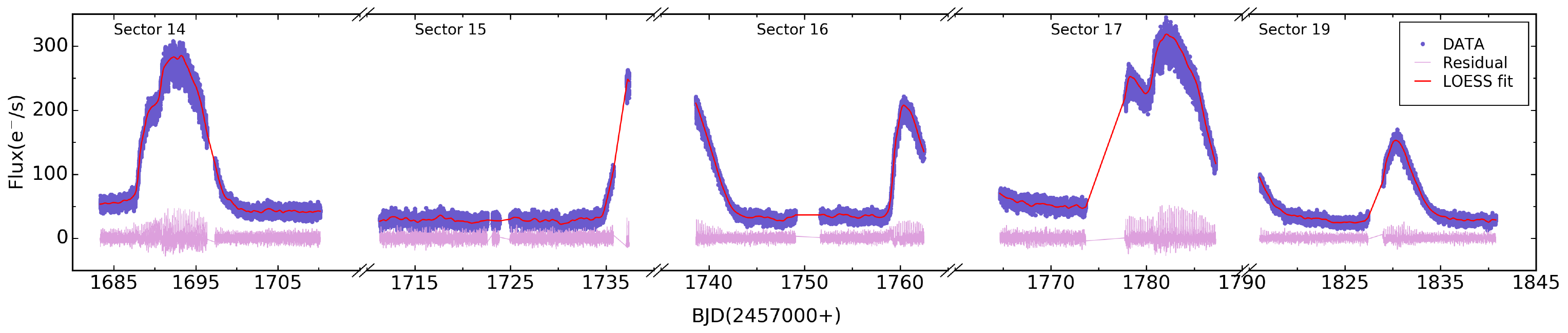}}
\subfigure[]{\includegraphics[width=0.92\textwidth]{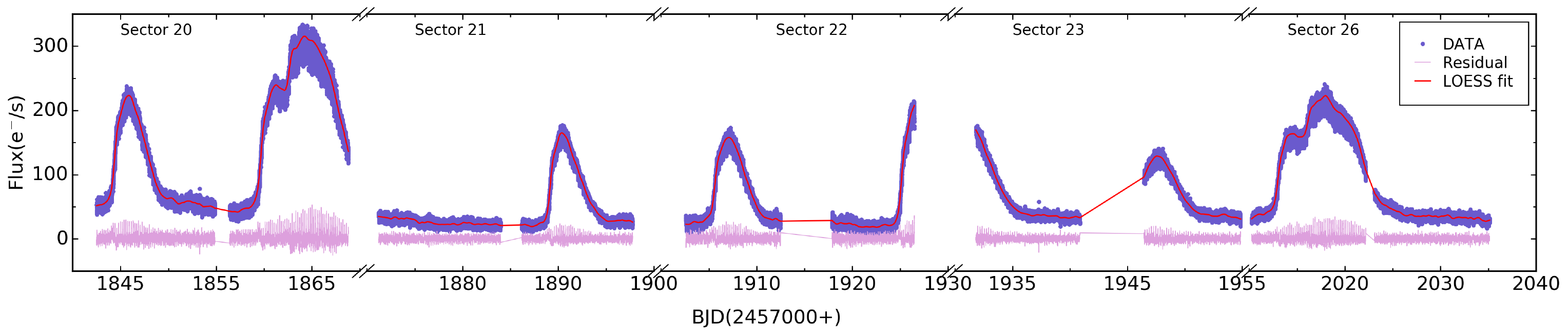}}
\subfigure[]{\includegraphics[width=0.92\textwidth]{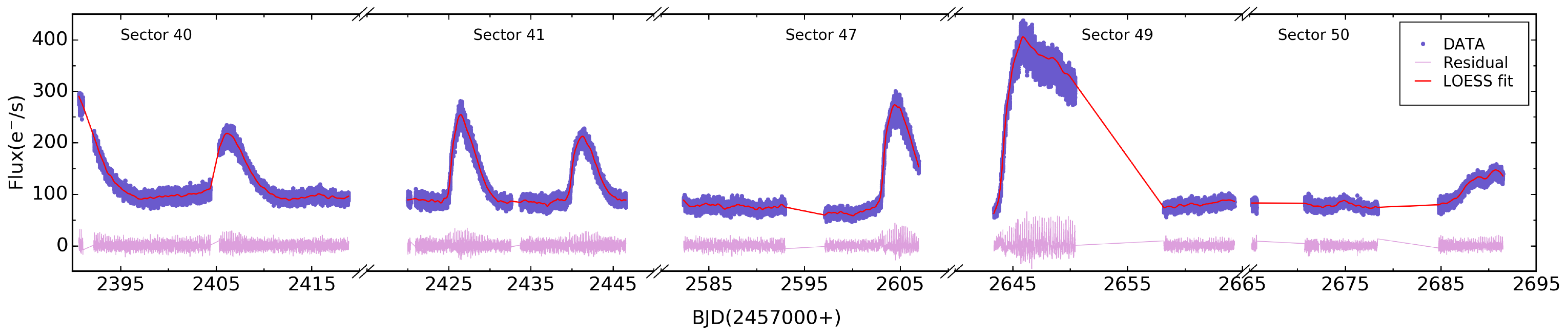}}
\subfigure[]{\includegraphics[width=0.92\textwidth]{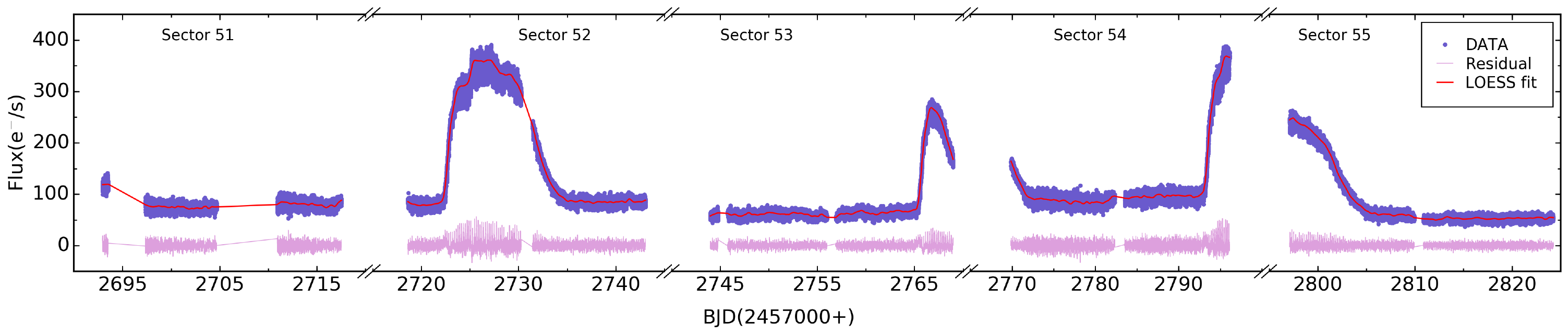}}
\subfigure[]{\includegraphics[width=0.92\textwidth]{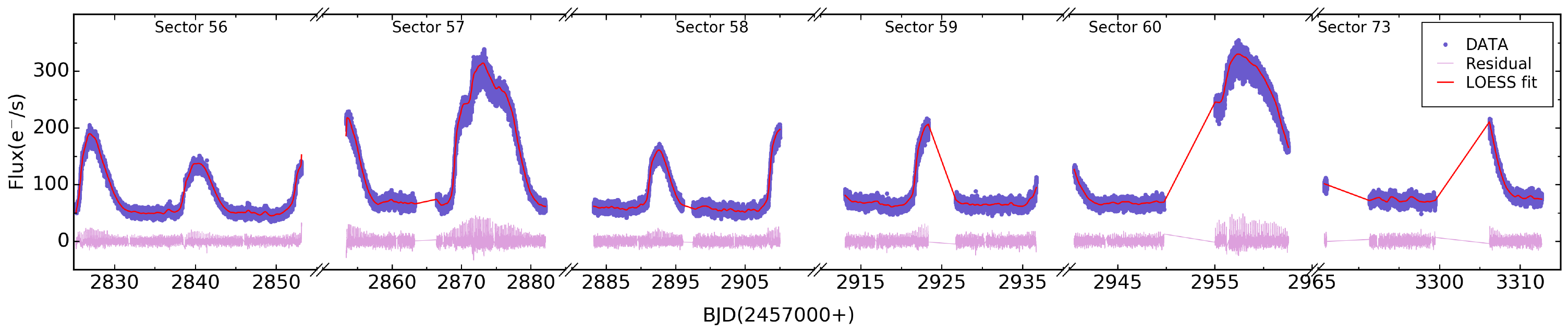}}
\caption{\textit{TESS} light curves of V416 Dra for all 26 sectors displaying outbursts and quiescence features. The red solid line represents the smoothed light curve using the LOESS fit. The light pink line is the detrended light curve by subtracting the smoothed light curve.}
\label{fig:tesslc_v416dra}
\end{figure*}

\begin{figure*}[ht]
\centering
\includegraphics[width=\textwidth]{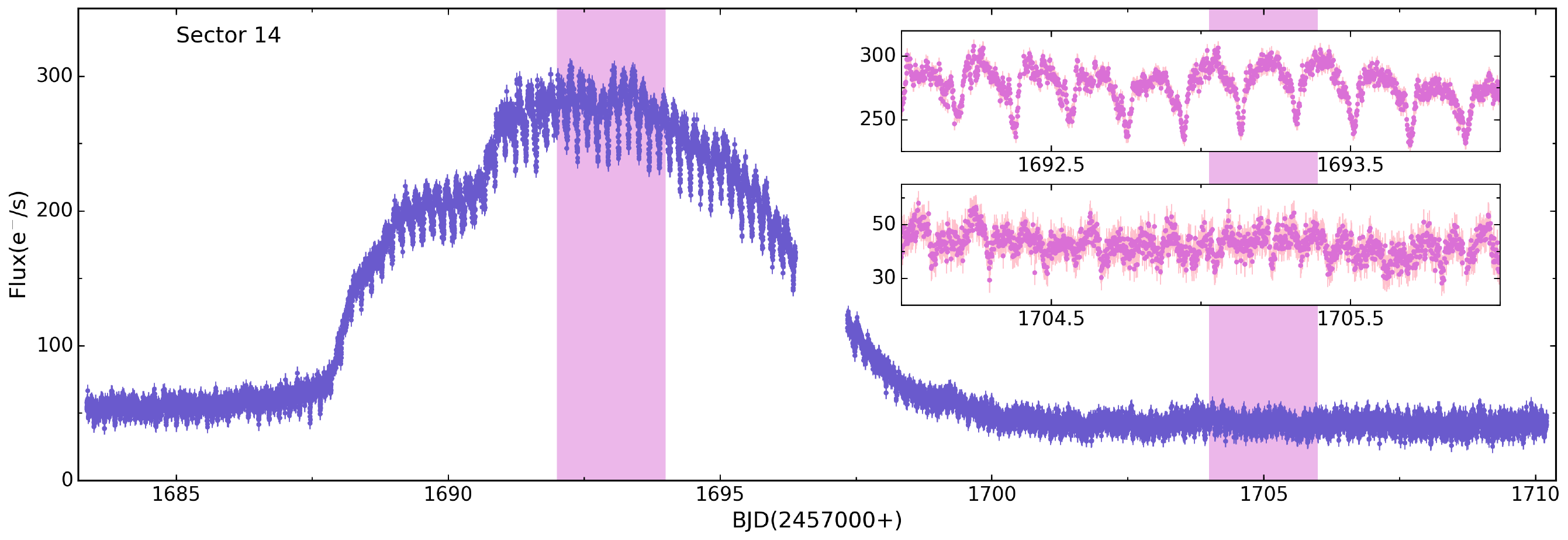}
\caption{The zoomed-in full \textit{TESS} light curve of V416 Dra for sector 14, with inset plots representing  the shaded-pink regions. These insets highlight the presence of eclipses during outbursts and quiescent durations.}
\label{fig:tesseclsec14_v416dra}
\end{figure*}

\begin{figure*}[h!]
\centering
\includegraphics[width=0.78\textwidth]{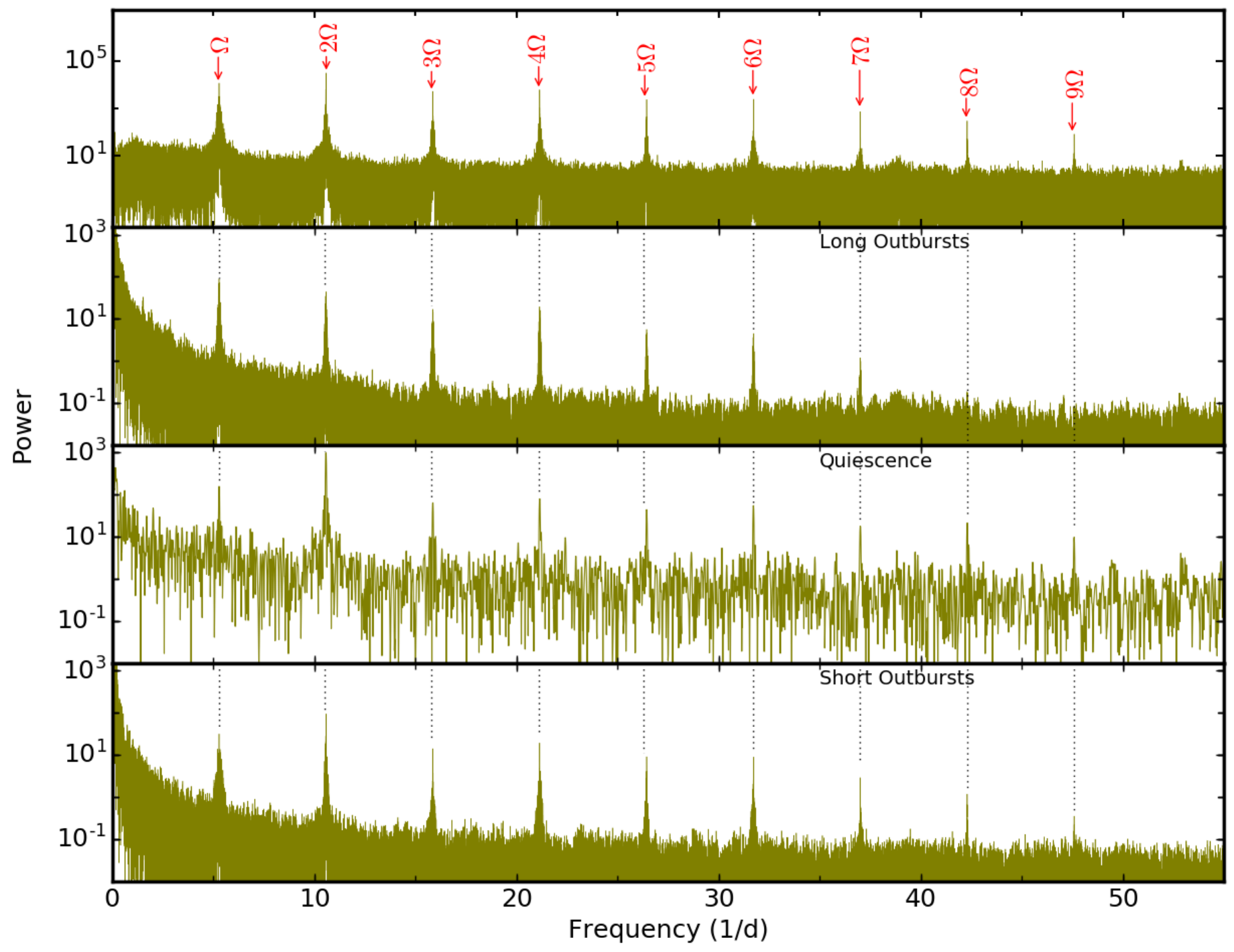}
\caption{Top panel represents the LS periodogram of V416 Dra obtained from the combined set of all sectors detrended light curves. The significant signals corresponding to the frequency $\Omega$, as well as its harmonics, are marked. The other three panels show the LS power spectra obtained by utilizing data during the long outburst, quiescent, and short outburst durations}. 
\label{fig:tess_ps_total_lo_qs_no_v416dra}
\end{figure*}

\begin{figure*}[h!]
\centering
\includegraphics[width=0.98\textwidth]{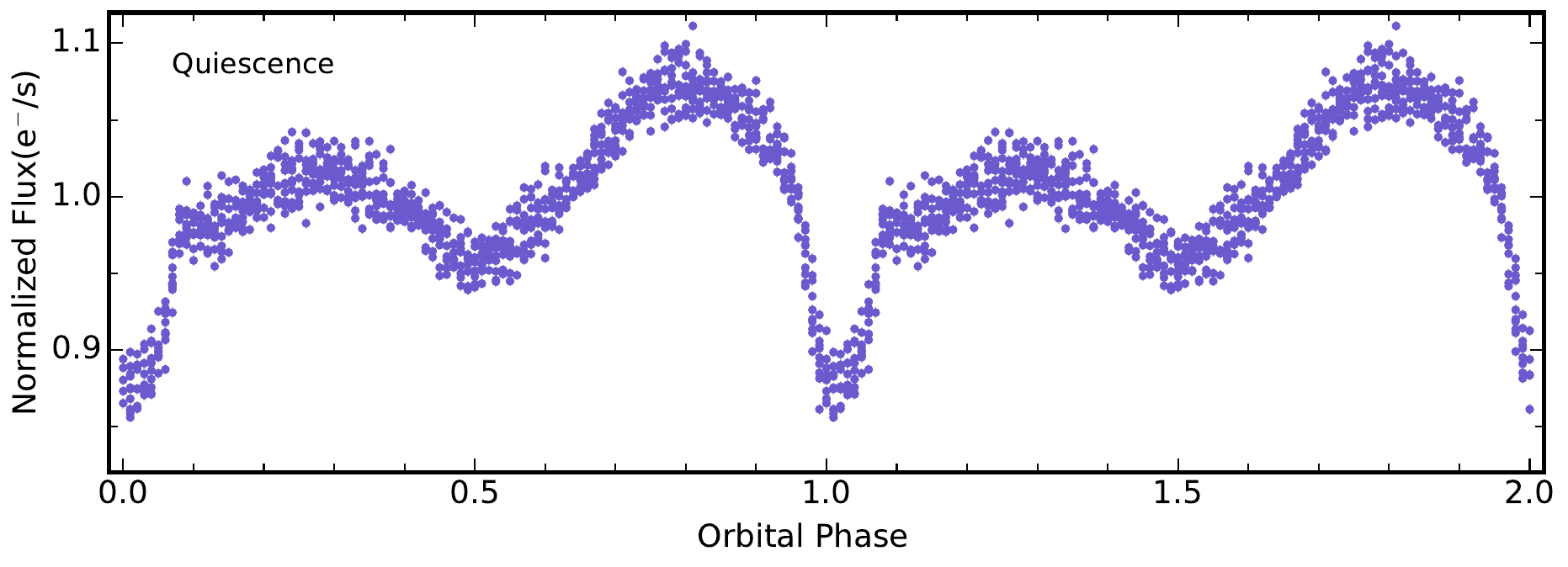}
\caption{Orbital-phase-folded light curve of V416 Dra obtained from the combined data of all sectors in the quiescent durations with phase bin of 0.001.} 
\label{fig:tessflc_qs_v416dra}
\end{figure*}

\begin{table*}[h!]
\small
\begin{center}
\caption{Identification, flux, EW, and FWHM for emission features in the spectra of J0808 and V416 Dra for the indicated measurement epochs. \label{tab:opt_spec_J0808_v416dra}}
\setlength{\tabcolsep}{0.05in}
\begin{tabular}{lcccccccccccccccccc}
 \hline
 &&&&&& \multicolumn{3}{c}{J0808} &&&&& \multicolumn{1}{c}{} \\
 \hline
  Identification &&\multicolumn{3}{c}{17-12-2018} &&\multicolumn{3}{c}{25-11-2022}&& \multicolumn{3}{c}{26-11-2022}\\
\hline 
                         &&Flux&$-$EW&FWHM &&Flux&$-$EW&FWHM &&Flux&$-$EW&FWHM \\
\hline
 H$\delta$(4102 \AA)   && 85.8$\pm$1.7  & 71$\pm$2  &4121$\pm$145  &&51.9$\pm$0.5  &57.9$\pm$1.9  &2633$\pm$17 && 49.2$\pm$0.3  & 52.5$\pm$0.5 & 2858$\pm$13    \\
 H$\gamma$(4340 \AA)   && 89.0$\pm$0.2  & 86$\pm$1  &3715$\pm$10    &&34.8$\pm$0.5  &31.8$\pm$0.6  &2001$\pm$14 && 29.4$\pm$1.1  & 26.2$\pm$1.4 & 2959$\pm$20    \\
 H$\beta$(4861 \AA)    && 57.4$\pm$0.6  & 52$\pm$1  &2157$\pm$13   &&48.5$\pm$0.4  &53.9$\pm$0.9  &1906$\pm$8  && 47.2$\pm$0.5  & 59.7$\pm$1.0 & 1982$\pm$11   \\
 HeI(5875 \AA)         && 8.4 $\pm$0.6  & 8 $\pm$1  &1525$\pm$40   &&12.8$\pm$0.1  &14.4$\pm$0.1  &1576$\pm$16 && 8.2$\pm$0.3   & 9.9$\pm$0.4  & 1950$\pm$38   \\
 H$\alpha$(6563 \AA)   && 51.1$\pm$0.9  & 53$\pm$1  &1437$\pm$10   &&74.5$\pm$0.1  &77.5$\pm$0.3  &1643$\pm$2  && 50.2$\pm$1.4  & 53.5$\pm$2.7 & 1548$\pm$26   \\
\hline
 &&&&&& \multicolumn{3}{c}{V416 Dra} &&&&& \multicolumn{1}{c}{} \\
 \hline
Identification       &&\multicolumn{3}{c}{21-03-2023}&&\multicolumn{3}{c}{21-04-2023} && \multicolumn{3}{c}{20-05-2023}\\
\hline
  &&Flux&$-$EW&FWHM  &&Flux&$-$EW&FWHM &&Flux&$-$EW&FWHM\\
\hline
 H$\delta$(4102 \AA) && 206.2$\pm$3.7  & 13.1$\pm$0.2 & 2336$\pm$15 && 358.4$\pm$8.0  &28.0$\pm$0.8&2523$\pm$32 && 244.4$\pm$10.8& 23.6$\pm$1.5 & 2412$\pm$36 \\
 H$\gamma$(4340 \AA) && 220.6$\pm$4.5  & 15.5$\pm$0.4 & 2256$\pm$24 && 290.2$\pm$3.9&22.9$\pm$0.4&1935$\pm$15 && 264.6$\pm$6.3 & 24.0$\pm$0.8 & 2267$\pm$29 \\
 H$\beta$(4861 \AA)  && 257.6$\pm$1.5  & 21.2$\pm$0.2 & 1840$\pm$6  && 360.2$\pm$4.6&32.3$\pm$0.6&1969$\pm$18 && 260.0$\pm$2.5 & 26.1$\pm$0.4 & 1914$\pm$8  \\
 HeI(5875 \AA)       && 22.8$\pm$1.0   & 2.3$\pm$0.1  & 629$\pm$19  && 30.4$\pm$1.4 &3.0$\pm$0.2 &643 $\pm$16  && 37.3$\pm$1.0 & 2.8$\pm$0.1  & 844$\pm$25 \\
 H$\alpha$(6563 \AA) && 295.2$\pm$3.7  & 33.3$\pm$0.6 & 1408$\pm$10 && 344.6$\pm$3.1&35.9$\pm$0.5&1420$\pm$8  && 475.3$\pm$4.4 & 36.7$\pm$0.5 & 1519$\pm$8 \\
\hline
\end{tabular}   
\end{center}
{Flux, EW, and FWHM are in units of 10$^{-16}$ erg cm$^{-2}$ s$^{-1}$, \AA, and km s$^{-1}$, respectively.}
\end{table*}


\section{Analysis and Results}\label{sec:analysis_results}
\subsection{CRTS J080846.2+313106}\label{sec:Analysis_J0808}
\subsubsection{Light-curve morphology and power spectra}
\label{sec:Analysis_phot}
Long-term light curves of J0808 were obtained from the CRTS, AAVSO, ZTF, and \textit{TESS} observations, which are shown in Figure \ref{fig:longtermlc_J0808}. The brightness of the system changes approximately in the range between  20 and 14 mag. The CRTS V light curve indicates an average V $\sim$ 19 mag and an outburst amplitude of $\sim$ 3$-$4 mag. The CRTS V dataset appears to detect thirteen outbursts; however, the source brightness drops from maximum light on timescales of approximately 5 and 8 days, according to AAVSO V (from JD 2456255.50 to JD 2456260.47) and CV (from JD 2459328.33 to JD 2459336.47) data, respectively. During ZTF observations, eleven outbursts appear to be visible, with one long outburst accompanying one small outburst. The short outbursts were observed approximately after 100 days of long outbursts. The shaded pink regions in Figure \ref{fig:longtermlc_J0808} corresponds to the \textit {TESS} observations. To probe J0808 in more detail, we have closely inspected its \textit{TESS} light curves independently. The full \textit{TESS} light curve is shown in Figure \ref{fig:tesslc_J0808}, which shows two types of outbursts that can be distinguished by the amplitude and duration of the outbursts. A long outburst was observed with a duration of $\sim$ 11.3 days in sector 44. Subsequently, the system is found to be in a quiescent state. Approximately 106 days after the first outburst, the system undergoes another one, this time with a duration of $\sim$ 2.5 days in sector 47. The amplitude of this outburst is somewhat smaller than the amplitude of the outburst observed in sector 44. After approximately 629 days, again this system was observed from \textit {TESS} for approximately 50 days, but no outburst was detected during this period.

Extensive \textit {TESS} observations  display statistically improved quiescence light curves, which offers a potential opportunity to measure its system parameters. Here, our first motive is to measure an orbital period of the system. To determine the orbital period of the system, the data must be detrended by removing the long-term trends, such as the outburst profile. Thus, we used locally weighted regression \citep[LOESS;][]{cleveland1979robust} with a smoothing span of 0.005 to smooth the light curves. Then, the smooth curves were subtracted from the original light curves to obtain the light curves with the outbursts removed. The LOESS fit and the detrended light curve are also shown in Figure \ref{fig:tesslc_J0808}. To search for the periodic signals in the light curve, we adopted the Lomb-Scargle \citep[LS;][]{Lomb76, Scargle82} periodogram algorithm to the combined set of all sectors detrended light curves, which is shown in Figure \ref{fig:tessps_J0808}. Two prominent peaks, corresponding to periods 4.9116$\pm$0.0003 h and 
   2.45586$\pm$0.00008 h, are observed in the power spectrum. We provisionally identify the derived period of $\sim$ 4.9116 h as the orbital period ($P_\Omega$) of J0808 and the $\sim$ 2.45586 h period as the second harmonic of the orbital period ($P_{2\Omega}$). We have also performed an LS periodogram analysis for the quiescent (from BJD 2459514.5319488 to BJD 2460285.5864469) and outburst (from BJD2459500.4000170 to BJD 2459511.7247343 in sector 44, and from BJD 2459604.0007540 to BJD 2459606.9451476 in sector 47) time-spans. The LS power spectrum obtained from the quiescent light curves is shown in the top panel of Figure \ref{fig:tessps_qs_ob_J0808}. Similar to the combined power spectrum, 
 two prominent signals at $P_\Omega$ and $P_{2\Omega}$ are observed in the quiescent power spectrum. However, only $P_\Omega$ is observed in the power spectrum obtained from the duration of the outburst phases (see bottom panel of Figure \ref{fig:tessps_qs_ob_J0808}). 
 
 To study the phased-light curve variations, we folded the \textit {TESS} light curves during both states using an arbitrary zero time BJD=2453480.776245071 (chosen so that the deeper minimum falls at phase 0.5), and our derived orbital period of 4.9116 h. Phase-folded light curves were obtained 
  for the duration of the quiescent and outburst phases with a phase bin of 0.05 along with a sum of two best-fit functions, a sine and a cosine, and are shown in Figure \ref{fig:tessflc_qs_ob_J0808}. The light curve pattern is distinct in both quiescence and eruption states. The quiescent profile exhibits double-peaked orbital modulations, with the maxima occurring near phases 0.2-0.4 and 0.7-0.9, and minima during conjunctions at phases 0 and 0.5. The peak in the light curve near phases 0.7-0.9 is higher than the peak near phases 0.2-0.4. The presence of a double-peaked pulse profile is also evident from the power spectrum, where along with orbital frequency a strong significant peak is observed at its second harmonic. However, a broad minimum appears filled with emission during the eruption phase.

\begin{table}
\centering
\caption{Periods corresponding to the dominant peaks in the power spectrum of V416 Dra obtained from the combined set of all sectors detrended light curves.}
\label{tab:ps_v416dra}
\renewcommand{\arraystretch}{1.4}
\begin{tabular}{lc}
\hline
Identification & Period (h)\\
\hline
\po      &  4.5386$\pm$0.0002    \\
\ptwoo   & 2.26931$\pm$0.00004   \\
\pthreeo & 1.51289$\pm$0.00002    \\
\pfouro  & 1.13466$\pm$0.00001     \\
\pfiveo  & 0.907730$\pm$0.000007   \\
\psixo   & 0.756441$\pm$0.000004     \\
\pseveno & 0.648377$\pm$0.000003    \\
\peighto & 0.567330$\pm$0.000003    \\
\pnineo  & 0.504293$\pm$0.000002    \\
\hline
\end{tabular}
\end{table}

\subsubsection{Optical spectroscopy}
\label{sec:Analysis_spec}
Optical spectra of J0808 exhibit a flat continuum and strong Balmer emission lines from H$\alpha$ to H$\delta$ during each epoch of observations. No broad Balmer absorption seems to be present in its spectrum. The observed wealth of spectral features resembles a typical spectrum of dwarf novae during quiescence. Figure \ref{fig:spec_J08} shows the optical spectra of J0808 for epochs 17 December 2018 and 25-26 November 2022. Identification, flux, equivalent width (EW), and FWHM of the principal emission lines of J0808 were obtained through a single-Gaussian fitting and are given in Table \ref{tab:opt_spec_J0808_v416dra}, where the error associated to each parameter corresponds to the standard deviation obtained from multiple spectral measurements with slight changes in the continuum fit. The flux ratio of HeI (5875\AA)/H$\alpha$ is measured to be approximately 0.2, which appears to be consistent with what is typically observed in dwarf novae \citep[e.g.,][]{Szkody81, Williams82, Thorstensen01, Breedt12}. Similar to \cite{Szkody04} and \cite{Han20}, strong emission lines are detected in the presented observations but with variable flux and EW values.

 To estimate the secondary star contribution, we have used K- and M-type dwarf spectra from the \cite{Pickles98} library. We scaled and subtracted them from our observed mean spectrum, resulting in acceptable cancellation of the late-type features that are commonly seen in M2-4V stars. This seems to agree with the expected M2.5 donor, as predicted by \cite{Knigge11} for the observed orbital period of  J0808.

\begin{figure*}
\centering
\includegraphics[width=0.3\linewidth]{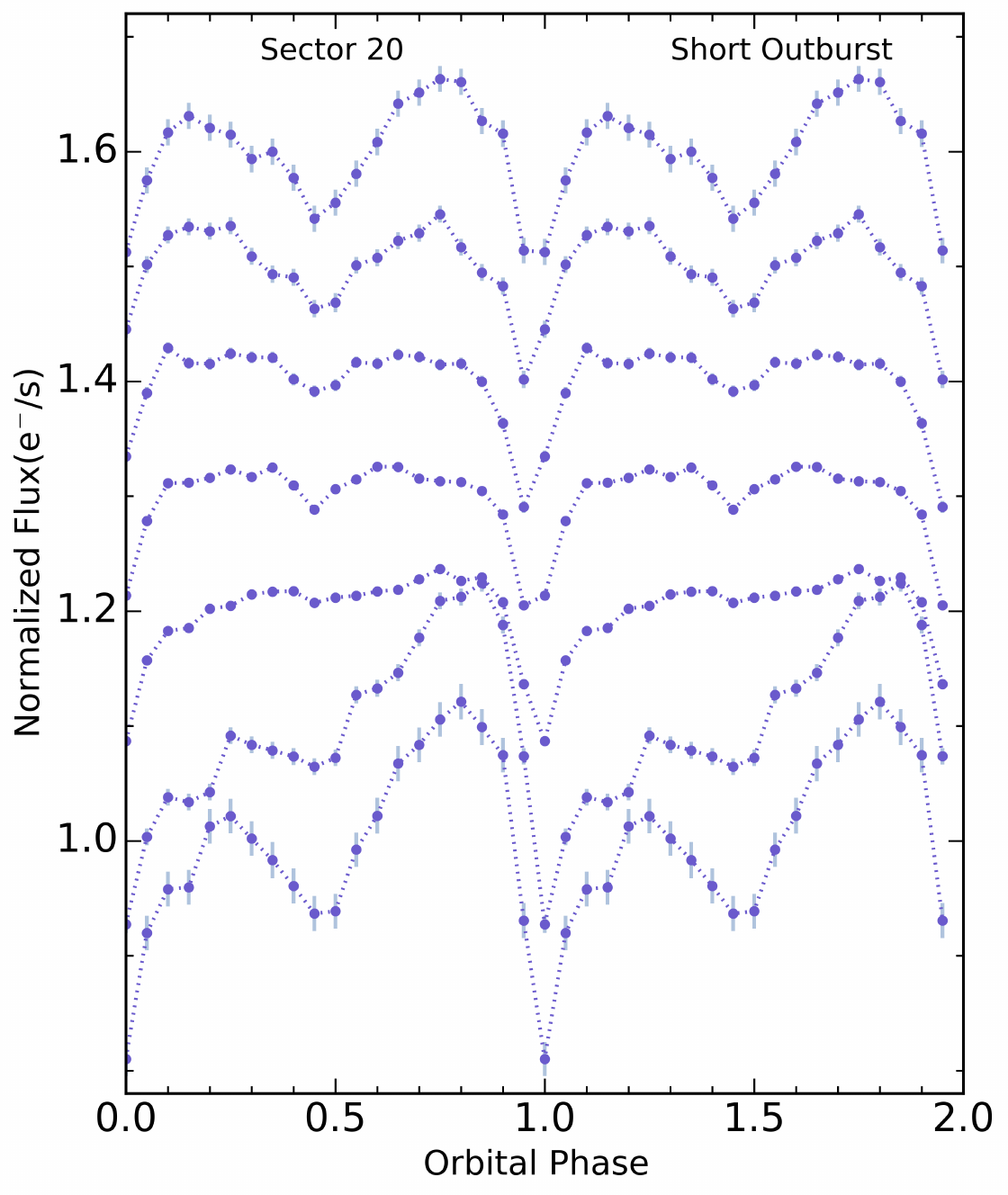}
\includegraphics[width=0.3\linewidth]{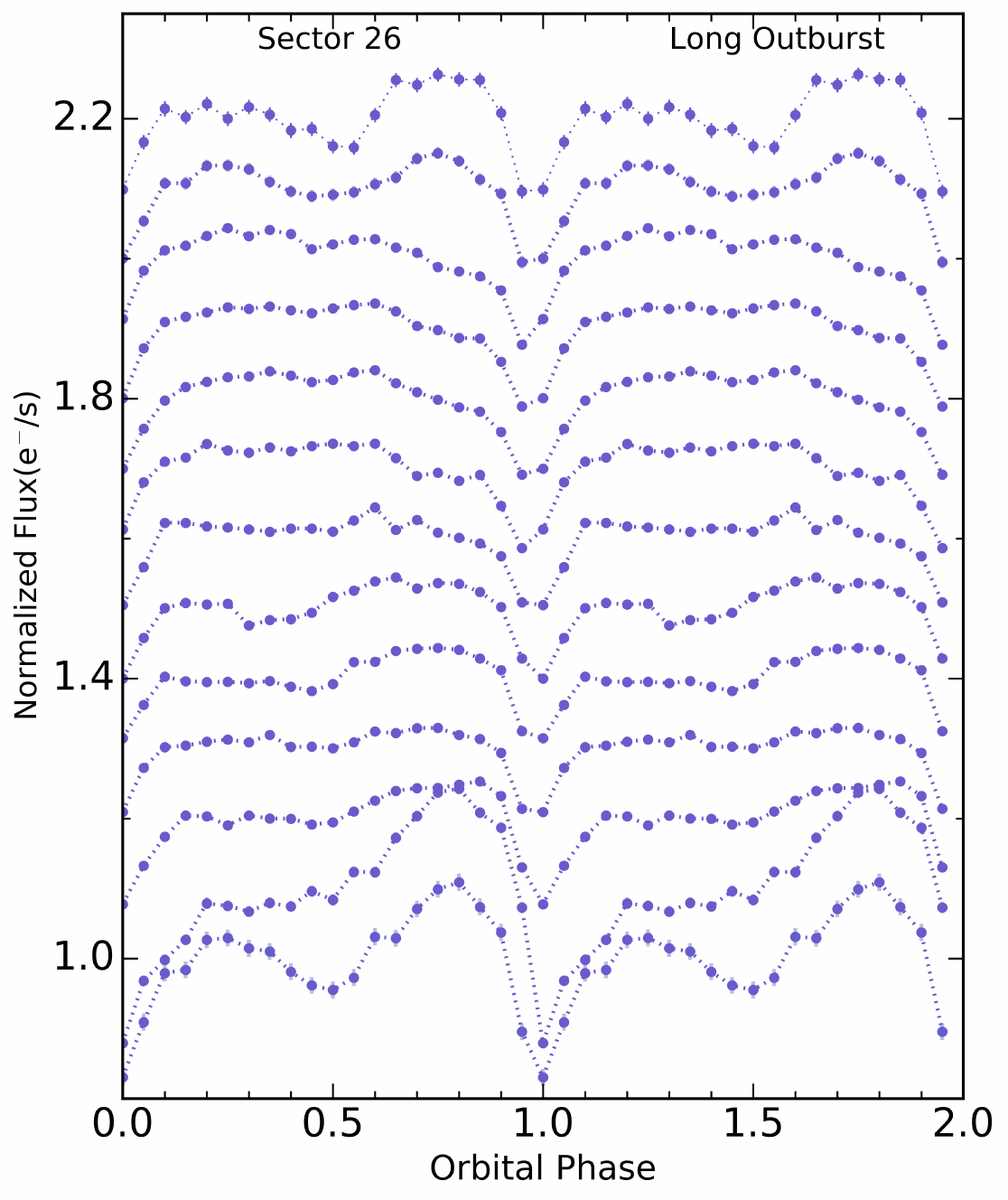}
\includegraphics[width=0.3\linewidth]{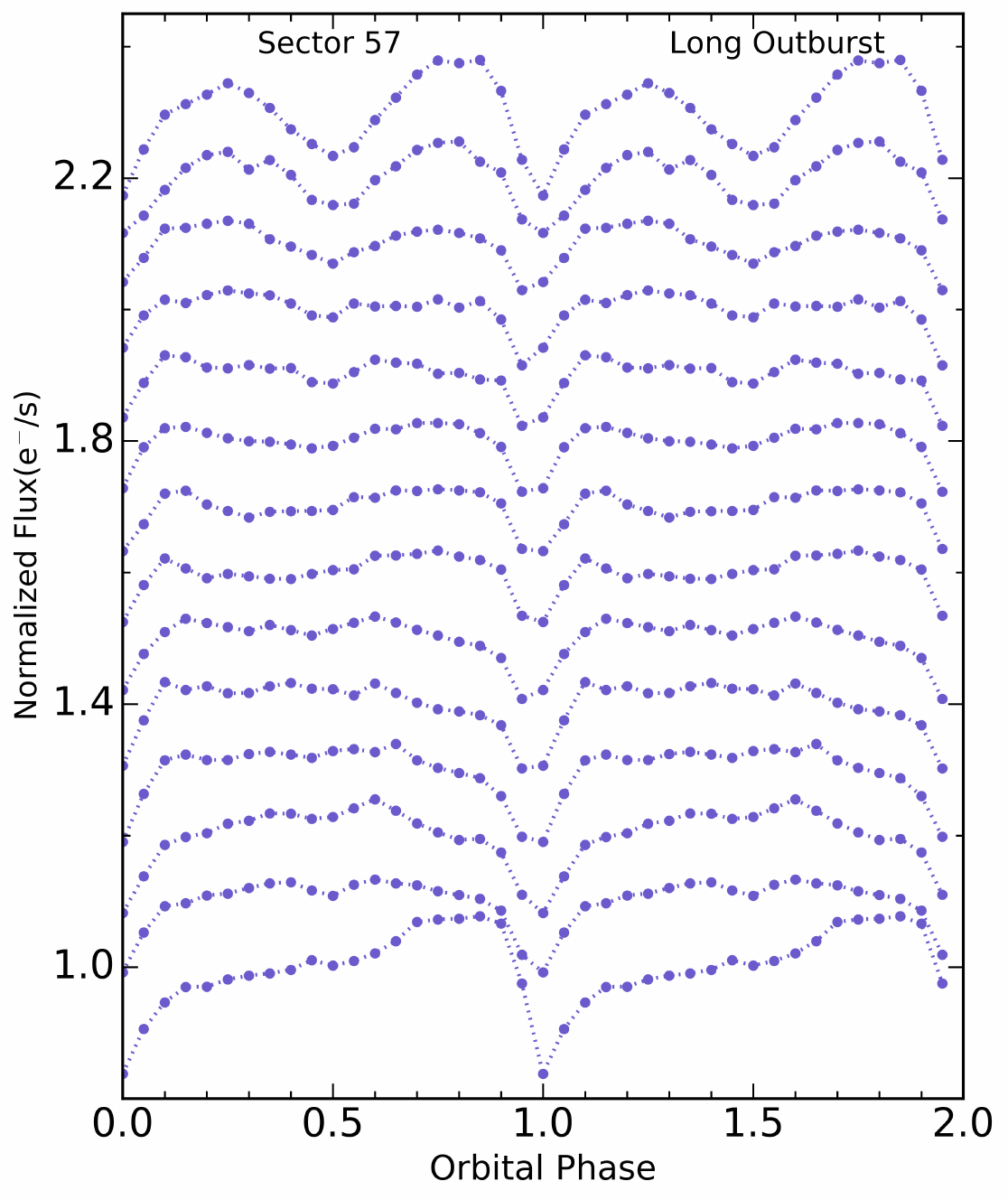}
\caption{Orbital-phase-folded light curves of V416 Dra obtained using the segmented five successive orbital cycles of the short and long outburst phases of sectors (from left to right) 20, 26, and 57, which are plotted in order of increasing time from the bottom up. For improved visibility, each successive light curve is vertically shifted by increments of 0.1 in the normalized flux.}
\label{fig:flc_sob_lob} 
\end{figure*}

\begin{figure*}
\centering
\includegraphics[width=0.9\textwidth]{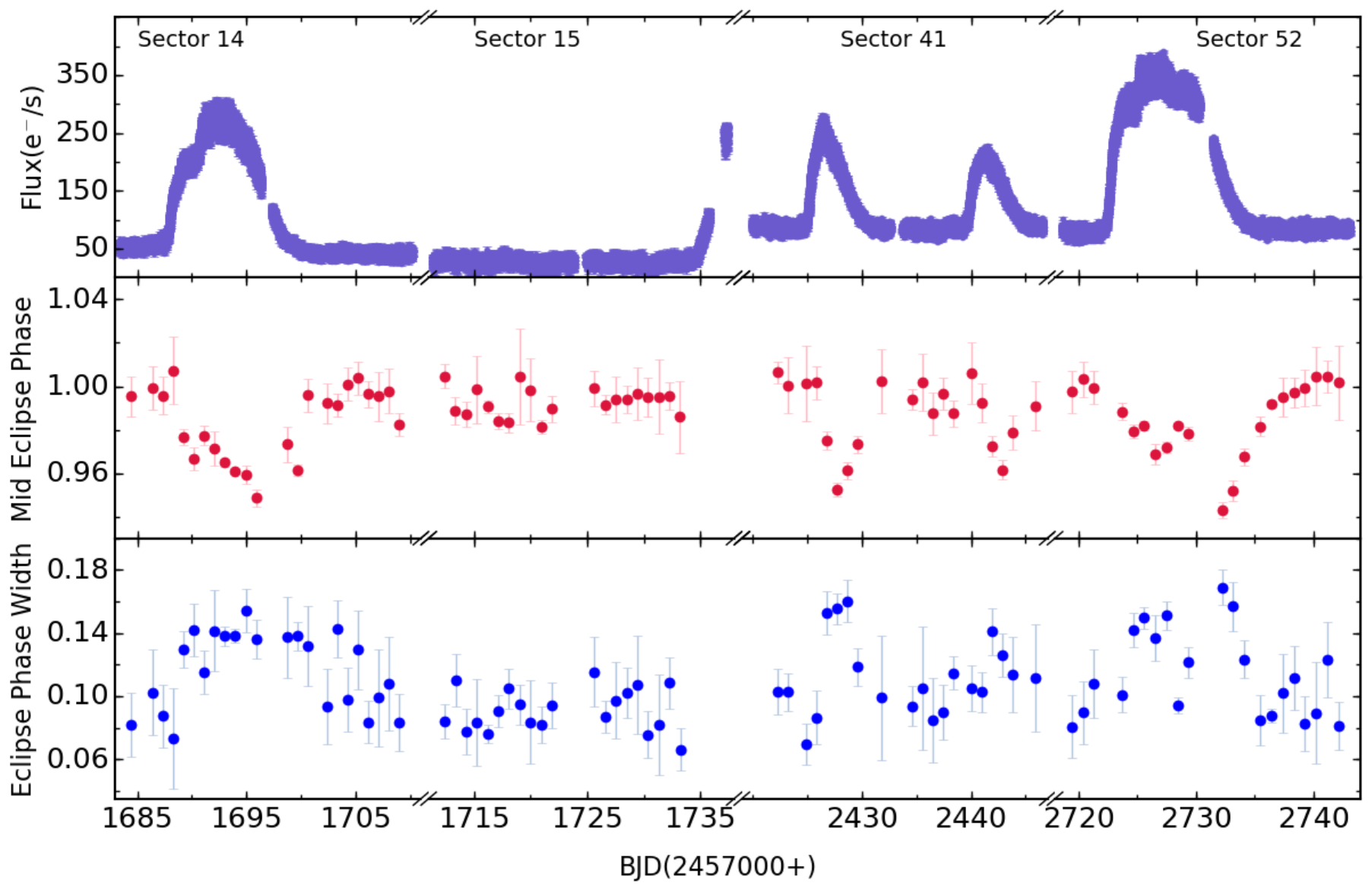}
\caption{Top panel represents the \textit {TESS} light curves of V416 Dra for sectors 14, 15, 41, and 52. The middle and lower panels represent the mid-eclipse phase and eclipse phase width, derived by fitting the mean eclipse-phased profile obtained from  five successive eclipses of these sectors light curves.}
\label{fig:lc_mid_ecl_phase_width_v416dra}
\end{figure*}

\begin{figure*}
\centering
\includegraphics[width=0.65\textwidth]{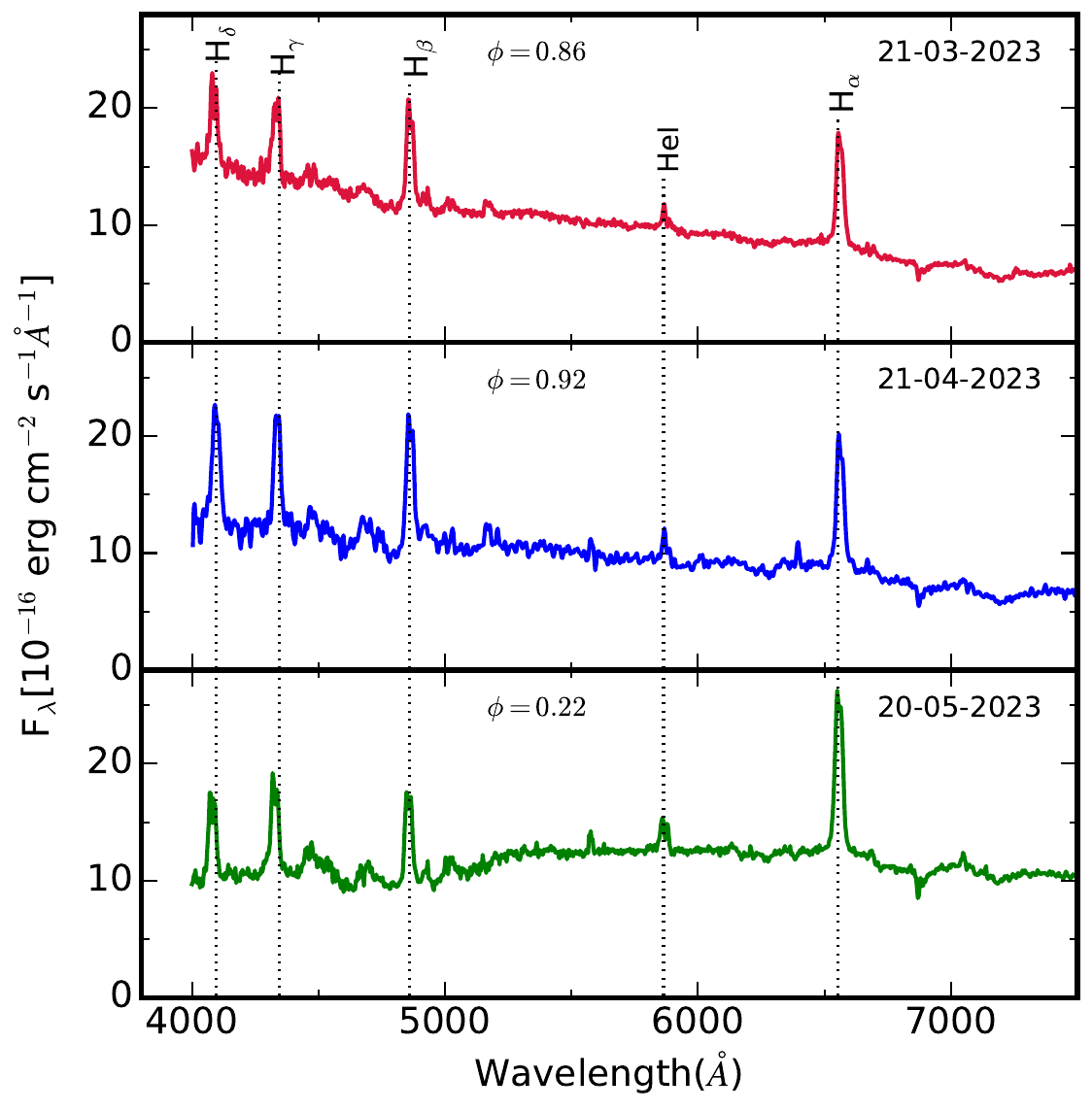}
\caption{Optical spectra of V416 Dra for three epochs of observations. The orbital phase and epoch of observation is mentioned in each panel.} 
\label{fig:spec_v416dra}
\end{figure*}
 

\subsection{V416 Dra}
\subsubsection{Light-curve morphology and  power spectra}\label{sec:Analysis_V416_Dra}
Figure \ref{fig:longtermlc_v416dra} displays the long-term light curve of V416 Dra that was obtained from the AAVSO, ZTF, and \textit {TESS} observations. \textit {TESS} observations are represented by the light blue shaded regions. The system brightness seems to vary between 18 and 13 mag. Multiple outbursts, including one large amplitude outburst followed by a series of much shorter or slightly fainter short outbursts, are observable in the long-term light curve. For \textit {TESS} observations, data spanning a total of 26 sectors are available to support the study of V416 Dra. The {\em TESS} light curves of V416 Dra for all 26 sectors are shown in Figure \ref{fig:tesslc_v416dra}, which reveals outburst as well as quiescent phase light curves. These light curves exhibit eclipses (see Figure \ref{fig:tesseclsec14_v416dra}). The \textit {TESS} light curves display a total of 30 outbursts, featuring two distinct types: broad and large-amplitude outbursts, as well as short outbursts with smaller amplitudes and sharp peaks. Among these, eleven are long outbursts, while the rest are short outbursts.  The overall shape of the long outbursts appears to vary slightly across observational epochs. After the initial rise, long outbursts in sectors 14, 17, 20, 26, and 60 either remain steady or decrease for approximately 1 to 1.5 days before increasing again. The long outburst in sectors 52 and 57, on the other hand, occurs in four stages: an initial rise followed by approximately one day of stability, then a subsequent increase followed by roughly three days of stability, a subsequent decline with another day of stability, and finally return to quiescence. All these  long-outbursts are observed with a duration of $\sim$ 13 days. However, short  outbursts occurred with a duration of $\sim$ 6 days but varying amplitudes. The time duration  between short outbursts, as well as between short and  long outbursts, is approximately 20 days.
  
 As discussed in Section \ref{sec:Analysis_phot}, the LS analysis was used to find the periodicity from the detrended light curves or the data from which the trend of outbursts was removed using a LOESS fit. The LS power spectrum obtained using the combined set of all sectors detrended light curves is shown in the top panel of Figure \ref{fig:tess_ps_total_lo_qs_no_v416dra}. The power spectrum contains 9 equidistant prominent peaks. The first peak, with a period of 4.5386$\pm$0.0002 h, agrees well with the period given by \cite{Aungwerojwit06}, which is the orbital period of the system. We also detected higher harmonics of the orbital frequency due to the non-sinusoidal light curve shape through each orbit, which are marked in the power spectrum and are given in Table \ref{tab:ps_v416dra}. 
  In addition, to search for the  superhumps, we have also performed LS periodogram analyses for the combined set of long outbursts, short outbursts, and quiescent phases separately, and obtained similar orbital frequency and their harmonics in each power spectrum (see Figure \ref{fig:tess_ps_total_lo_qs_no_v416dra}). Moreover, we  have also segmented the data into sections corresponding to individual long-outbursts, quiescent, and short outburst phases across all sectors, and performed power spectral analysis for each section. Nevertheless, no other periodic pattern was observed near the orbital frequency region during these time-spans.

\subsubsection{The Eclipse Profile}
We explored the phased light-curve variation of V416 Dra during its binary motion. Light curves obtained from the combined \textit {TESS} observations during quiescent phases are folded using the zero epoch given by \cite{Aungwerojwit06} and our derived \textit {TESS} orbital period. The folded light curve derived for the duration of the quiescent phases with bins of width 0.001 is shown in Figure \ref{fig:tessflc_qs_v416dra}. Most notably, there is an increase in fluxes during quiescence at phases 0.2-0.4 and 0.7-0.9, and the minimum at 0.0 is deeper than the minimum at 0.5. A bright hump feature preceding the eclipse was also observed by \cite{Aungwerojwit06} in the light curves of V416 Dra.

To observe the characteristics of the eclipse profile for the duration of the short-outburst and long-outburst phases, we divided each light curve into sections comprising five successive orbital cycles within these phases. Similar to the approach as described above for quiescence, the folded light curves corresponding to these phases were extracted with bins of width 0.05. Similar to \cite{Aungwerojwit06}, the eclipse depth and profile appear to have changed from quiescence to eruption. We have found that as the brightness increases during outbursts in each light curve, the brightness outside the eclipse becomes much less pronounced or remains relatively flat, and the eclipse feature is V-shaped and symmetrical, as is usual for an eclipsing dwarf nova in outburst \citep[see][]{Baptista2000}. Additionally, peaks emerge at orbital phases 0.2-0.4 and 0.7-0.9 in the folded light curve as the system approaches quiescence. For example, Figure \ref{fig:flc_sob_lob} shows the phased light curves corresponding to short and long-outbursts obtained from sectors 20, 26, and 57, which clearly demonstrate that over the course of the outburst, brightness becomes more uniform. 

Subsequently, the phase-folded and binned eclipse profiles obtained from the segmented five successive orbital cycles during both outburst and quiescent episodes were fitted using a Gaussian model, which allowed us to determine the mid-eclipse phase and eclipse phase width for each sector data. For example, Figure \ref{fig:lc_mid_ecl_phase_width_v416dra} shows the mid-eclipse phase and eclipse phase width for \textit{TESS} light curves of sectors 14, 15, 41, and 52, which include short, long, and quiescent 
 episodes. We have found that the phase of the mid-eclipse is centred near 0.0 during quiescence, whereas the phase of the mid-eclipse occurs earlier during outbursts (short and long) compared to quiescence. Also, during outbursts, the full width at half-maximum of the eclipse is greater compared to quiescence. The average eclipse width at half  maximum ($\Delta \phi$) during quiescence and outburst is 0.082$\pm$0.005 and 0.124$\pm$0.009, respectively, implying that the eclipse duration in quiescence is less than that in outburst. 

Furthermore, we derive the orbital inclination ($i$) of the binary system by using the equation given by \cite{Eggleton83},
\begin{equation}
\left( \frac{R_2}{a} \right)^2  = \sin^2(\pi \Delta \phi) + \cos^2(\pi \Delta \phi) \cos^2i,
\end{equation}
where $R_2$ is the volume radius of the secondary star and $a$ is the binary separation. $R_2/a$  depends only on the mass ratio ($q$) as 
\begin{equation}
\frac{R_2}{a} =  \frac{Cq^{2/3}}{Dq^{2/3}+\ln(1+q^{1/3})}.
\label{eq:R2/a}
\end{equation}  
\noindent
The coefficients $C$ and $D$ are given by \citet{Eggleton83} as 0.49 and 0.6, respectively, for a spherical Roche lobe. However, in the case of CVs, the Roche lobe of the secondary is not spherical, having the largest size as ``seen'' from the WD and the smallest in the polar direction. Therefore, we have used the coefficients $C=0.4990$ and $D=0.5053$, as described by \cite{Andronov14}, assuming a significantly more accurate model for its elliptical projection onto the celestial sphere. Adopting a $\Delta \phi$ of 0.082 and a mass ratio $q$ of 0.5, 
 assuming an average mass of the WD in CVs of approximately 0.8 $M_\odot$ \citep{Zorotovic11, Pala20} and an expected secondary mass of approximately 0.4 $M_{\odot}$ \citep {Knigge11}, the orbital inclination is estimated to be $i$ $\sim$ 76\hbox {$^\circ$}. 

We have also performed an analysis of the accretion disc eclipses by comparing the circularization radius of the disc to the donor radius as reported by \cite{Court20}. The circularization radius can be calculated using the equation given by \cite{Frank02},
\begin{equation}
\frac{R_{\rm circ}}{a}   = (1+ q) (0.5-0.277\log q)^4.
\label{eq:Rcirc}
\end{equation}
 Adopting an expected average mass ratio $q$ of 0.5, the Roche lobe donor radius ($R_{2}$) and circularization radius ($R_{\rm circ}$) of the disc are $0.35a$ and $0.17a$, respectively. Additionally, we have estimated the size of the accretion disc and the half chord of the donor by combining the expected values for $q$, $i$, and $\Delta \phi$ with equations (4) and (5) of \cite{Dhillon91}. Using $\Delta \phi$ $\sim$ 0.124 during outburst, $q$=0.5, and $i$=76\hbox {$^\circ$}, we have estimated a disc radius and half chord radius as $0.43a$ and $0.25a$, respectively.

\subsubsection{Optical spectroscopy}
Figure \ref{fig:spec_v416dra} displays strong hydrogen Balmer emission lines from H$\alpha$ to H$\delta$, as well as HeI (5875 \AA) emission line in the optical spectra obtained during three distinct epoch observations of V416 Dra. The flux ratio of HeI (5875\AA)/H$\alpha$ is measured to be around 0.1. The observed optical spectra resemble the spectrum obtained by \cite{Aungwerojwit06}. Similar to \cite{Aungwerojwit06}, double-peaked Balmer emission lines are observed in all observed spectra. Identification, flux, EW, and FWHM of the principal emission lines of V416 Dra are given in Table \ref{tab:opt_spec_J0808_v416dra}. Similar to J0808, the comparison with the \cite{Pickles98} template spectra suggests that V416 Dra also exhibits a secondary of spectral type M2-4V, which seems consistent with the spectral type of M2.9  predicted by \cite{Knigge11} for the observed $P_\Omega$=4.5386 h.


\section{Discussion}
\label{sec:dis}
\subsection{J0808}
The observed long outburst following the short outburst after a long hiatus reveals J0808 to be a U Gem-type dwarf nova \citep[see][]{VanParadijs83, Szkody04}. A photometric period of 4.9116$\pm$0.0003 h is detected from the \textit {TESS} observations, which we provisionally suggest as the orbital period of this system. This derived period differs from U~Gem's (the class prototype) orbital period by only $\sim$ 0.66 h, lending further credence to J0808 as showing U Gem-type characteristics.

Furthermore, J0808 displays a double-peaked orbital pulse profile during quiescence, with unequal minima and maxima. The observed double-humped features suggest that these variations are the result of an ellipsoidal modulation of the secondary \citep{Warner95}. This is further evident from the detection of the dominant power at $2\Omega$, which reinforces the strong contribution from the secondary star due to its ellipsoidal modulation. The different sizes of maxima and minima features in J0808 suggest the existence of a second light source or an additional contribution from the accretion stream or hot spot in the system. The two maxima are the result of maximum visibility of the Roche-deformed sides of the secondary star at orbital phases 0.2-0.4 and 0.7-0.9, where an additional contribution from the accretion stream or hot spot could be partially obscured and unobscured, respectively. The shallow and deeper minima represent the inferior and superior conjunctions of the secondary. At superior conjunction, the most probable obstruction between the observer and the combined light of the secondary and the additional source is a faint accretion
disc. The variability shape is different during outburst as compared to quiescence. The deeper minimum phase of quiescence seems filled in with emission during the outburst, while the shallower minimum phase of quiescence becomes deeper and wider. The deeper minimum phase during the outburst is likely caused by the secondary eclipsing the component of the system from which the outburst originates.

 \subsection{V416 Dra}
The presence of long outbursts, along with frequent short outbursts, suggests that V416 Dra also belongs to the category of U Gem-type dwarf novae. The short outbursts exhibit a rapid rise followed by an exponential type decay. However, most of the long outbursts show an initial increase in brightness (similar to the short outbursts) followed by a plateau lasting less than one and a half days. Subsequently, there is a further rise in brightness, followed by a gradual decrease in flux (see Figure \ref{fig:tesslc_v416dra}). This observed pattern of behaviour is commonly referred to as a `precursor', which has been consistently observed in all superoutbursts of SU UMa-type CVs \citep[e.g.,][]{Still10, Barclay12}. However, after the emergence of high cadence data, \cite{Cannizzo12} and \cite{Ramsay12} suggested that these precursors are common to long outbursts from accreting CVs in general. The observed embedded precursors in V416 Dra support the findings of \cite{Cannizzo12} and \cite{Ramsay12} that long outbursts in dwarf novae above the period gap appear to be triggered by short outbursts. In this context, when the material in the outer radii of the accretion disc becomes ionised, a normal (or short) outburst is thought to generate a super- (or long) outburst. The long-baseline \textit{TESS} observations allow the orbital period to be unambiguously determined as 4.5386$\pm$0.0002 h, which is well consistent with the previously reported value. The observed orbital period, along with the lack of detection of any superhump period during long outbursts,  resembles the characteristics of U Gem-type dwarf novae.

The orbital modulation of V416 Dra reveals eclipse features, with noticeable variations in eclipse profiles when the system switches between outburst and quiescence. Similar to J0808, a characteristic of ellipsoidal variation of the secondary is also evident in the observed light curve of V416 Dra during quiescence, with an additional contribution from the accretion stream or hot spot. The dominant $2\Omega$ signal also strengthens the evidence for ellipsoidal modulation of the secondary. The deeper minimum during quiescence exhibits the occultation of the white dwarf and bright hotspot. The increase in flux just preceding the eclipse at orbital phase 0.7$-$0.9 is associated with the hotspot where the accretion stream from the donor star hits the accretion disc and is seen approximately face-on \citep[see][]{Wood86}. A similar pattern is observed in other eclipsing cataclysmic variables like V447 Lyr, KIS J192748.53+444724.5, Z Cha, and IPHAS
J051814.34+294113.2 \citep{Ramsay12, Scaringi13, Court19, Han21}. In these systems, the disc becomes larger and hotter during outburst, resulting in deeper eclipse depths compared to out-of-eclipse flux. Unlike in quiescence, the flat out-of-eclipse feature and V-shaped eclipse observed during V416 Dra outburst suggest that the optical light is dominated by the accretion disc. Consequently, the secondary star becomes less visible, and the bright spot makes a significantly smaller contribution to the total optical flux as compared to quiescence, and/or the hotspot emitting region has changed from a small, compact region to a larger structure over the disc. This is also evident from the derived mid-eclipse phase and width of the eclipse during outburst and quiescence episodes. The phase of the mid-eclipse occurs earlier during outbursts compared to quiescence, and the width of the eclipse is greater during an outburst. This further suggests that the accretion disc has a larger radius during outburst compared to during quiescence. Moreover, the Roche lobe-filling donor radius is estimated to be larger than the circularization radius of the disc, indicating that the accretion disc in quiescence is likely to be completely obscured. To investigate the possibility of a total eclipse of the disc during the outburst, we have also estimated the size of the accretion disc and the half chord and found that the disc radius during the outburst is larger than the half chord of the donor, implying that the donor star covers a significant fraction of the disc during the outburst. This also suggests that the disc grows in size as the outburst progresses, driven by the presence of high-viscosity material associated with the outburst. Conversely, the disc radius contracts during quiescence due to the accretion of low angular momentum material \citep{Smak71, Lasota01}.


\section{Summary} \label{sec:sum}
We have carried out detailed analyses of J0808 and V416 Dra, using optical photometry and spectroscopy. The observed outbursts and the absolute G-band magnitude of approximately +7.4 reveal their typical characteristics of non-magnetic CVs \citep{Warner87}. The light curve behavior, orbital periods, and spectra of both systems show remarkable similarity and appear to fall into the category of U Gem-type systems. Additionally, they appear to mirror the characteristics of the rare type dwarf nova CW 1045+525 \citep{Tappert01}. To summarize, we find the following characteristics of J0808 and V416 Dra: 
\begin{enumerate}
\item For the first time, an unambiguous period of 4.9116$\pm$0.0003 h is detected for J0808, which appears to be the probable orbital period of the system. This orbital period places it above the period gap in the distribution of orbital period of CVs. 
\item In both long-period systems, the observed dominant $2\Omega$ signal and the orbital modulation during quiescence suggests ellipsoidal modulation from the changing aspect of the secondary, with an additional contribution from the accretion stream or hot spot. However, during the outburst, the hotspot itself is overwhelmed by the increased brightness, which is possibly associated with the accretion disc.
\item The mid-eclipse phase for V416 Dra starts earlier and the eclipse width widens during outbursts compared to quiescence, implying a larger accretion disc radius during outbursts. This is supported by the accretion disc eclipse study of V416 Dra, suggesting a total disc eclipse in quiescence, while a significant fraction of the disc appears to be eclipsed during outbursts, indicating a possible growth in disc size during the outburst phase.
\item Optical spectra of J0808 and V416 Dra are typical of dwarf novae during quiescence. The lack of HeII (4686 \AA)  in the spectra of both systems rule out the possibility of magnetic CVs and nova-like variables or any hot optically thick discs. Additionally, both stars  exhibit a significant contribution from a M2-4V secondary.
\item The photometric  and spectroscopic  results obtained from both systems are very similar and can be placed within the classification of U Gem-type systems.
\end{enumerate}


\section*{Acknowledgements}
We thank the anonymous referee for the careful reading of the manuscript and giving us constructive comments and suggestions that improved the manuscript considerably. AJ acknowledges support from the Centro de Astrofisica y Tecnologias Afines (CATA) fellowship via grant Agencia Nacional de Investigacion Desarrollo (ANID), BASAL FB210003. 
 Support for MC is provided by ANID's FONDECYT Regular grant \#1171273; ANID's Millennium Science Initiative through grants ICN12\textunderscore 009 and AIM23-0001, awarded to the Millennium Institute of Astrophysics (MAS); and ANID's Basal project FB210003.
 This research includes data collected with the \textit{TESS} mission, obtained from the MAST data archive at the Space Telescope Science Institute (STScI). Funding for the \textit {TESS} mission is provided by the NASA Explorer Program. This research is also based on observations obtained with the Samuel Oschin 48 inch telescope at the Palomar Observatory as part of the Zwicky Transient Facility project. ZTF is supported by the National Science Foundation under grant No. AST-1440341 and a collaboration including Caltech, IPAC, the Weizmann Institute for Science, the Oskar Klein Center at Stockholm University, the University of Maryland, the University of Washington, Deutsches Elektronen-Synchrotron and Humboldt University, Los Alamos National Laboratories, the TANGO Consortium of Taiwan, the University of Wisconsin at Milwaukee, and Lawrence Berkeley National Laboratories. Operations are conducted by the COO, IPAC, and UW. We acknowledge with thanks the variable star observations from the AAVSO International Database contributed by observers worldwide and used in this research. The CRTS survey is supported by the U.S. National Science Foundation under grants AST-0909182 and AST-1313422. AJ acknowledges J. Joshi for reading our work and providing helpful comments.
  RD acknowledges funds by ANID grant FONDECYT Post-doctorado N$^\circ$ 3220449. The observing staff and observing assistants of 2.01-m HCT telescope are deeply acknowledged for their support during optical spectroscopic observations.


\bibliographystyle{aa}
\bibliography{ref}

\end{document}